\documentclass[journal]{IEEEtran}
\usepackage{tikz}
\usetikzlibrary{shapes.geometric, arrows, positioning, fit, backgrounds, calc}
\usepackage{adjustbox}
\usepackage{amsfonts}
\usepackage{amsmath}
\usepackage[ruled,linesnumbered]{algorithm2e}
\usepackage{array}
\usepackage{bm} 
\usepackage{booktabs}
\usepackage{color}
\usepackage{float}
\usepackage{graphicx}
\usepackage{indentfirst}
\usepackage{longtable}
\usepackage{multirow}
\usepackage{pifont}
\usepackage[caption=false,font=footnotesize]{subfig}
\usepackage{tabularx}
\usepackage{threeparttable}
\usepackage{wasysym}
\usepackage{yhmath}
\usepackage{newclude}
\usepackage{cite}
\usepackage{hyperref}
\usepackage{amssymb}

\title{Tri-Domain Multiuser MIMO Precoding Optimization and Channel Estimation with Spatial-EM Reconfigurable Antenna}

\author{Yining Li, Ziwei Wan, Zhen Gao,~\IEEEmembership{Senior Member,~IEEE}, Keke Ying, Lipeng Zhu,~\IEEEmembership{Member,~IEEE}, and Rui Zhang,~\IEEEmembership{Fellow,~IEEE}}

\begin{document}
\maketitle

\begin{abstract}
In this paper, we propose a tri-domain reconfigurable multiuser multiple-input multiple-output (MIMO) communication system that integrates the electromagnetic (EM) reconfigurable antenna (EMRA) with the spatially movable antenna (SMA), termed the spatial-EM reconfigurable antenna (SEMRA). The proposed system offers EM, spatial, and digital domain degrees of freedom (DoFs) for joint channel reconfiguration, yet introduces new challenges in channel estimation (CE) and precoding optimization. Specifically, for a multiuser orthogonal frequency division multiplexing (OFDM) downlink, the precoding design is formulated as a tri-domain optimization problem over antenna positions, EM-domain radiation-pattern weights, and digital precoders. We first develop a zero-forcing (ZF)-based baseline algorithm to decouple the design of spatial reconfiguration, and then propose a weighted minimum mean square error (WMMSE)-based tri-domain joint optimization algorithm to further improve the spectral efficiency (SE). Furthermore, we propose a low-overhead movement-aided channel estimation scheme in which coordinated antenna repositioning across pilot slots synthesizes a denser virtual array, enabling more accurate angle-of-departure (AoD) estimation and EM-domain channel state information (eCSI) reconstruction under the same per-user pilot overhead as the EMRA baseline. The resulting parametric representation enables eCSI assembly at desired antenna positions without additional pilots. Simulation results show that the proposed CE scheme improves eCSI estimation accuracy and the proposed SEMRA achieves higher SE than the EMRA baseline under the same pilot overhead.
\end{abstract}

\begin{IEEEkeywords}
Spatial-EM reconfigurable antenna, channel estimation, precoding, spectral efficiency, tri-domain hybrid precoding, weighted minimum mean square error.
\end{IEEEkeywords}

\section{Introduction}\label{S1}

Future wideband multiuser multiple-input multiple-output (MIMO) systems extend from terrestrial deployments to low Earth orbit satellite constellations and broader near-space integrated networks~\cite{Ying2023JSACQRA,Liu2025SpaceNSComNet,Liu2024SpaceNSCom,Li2026CJEEmbodied}. These systems require higher spectral efficiency (SE) under increasingly tight hardware constraints. In traditional fixed-antenna (TFA) settings, performance is improved mainly through digital precoding and array scaling, whereas antenna locations and radiation patterns remain largely fixed after fabrication. This limitation has motivated growing interest in flexible hardware architectures that introduce new physical-layer degrees of freedom (DoFs). Active transceiver architectures~\cite{Krikidis2024TCOMRAMIMO,Wong2022TWCFAMA}, such as MIMO systems with electromagnetic reconfigurable antenna (EMRA), have shown substantial SE improvement without extra antennas or power-hungry radio-frequency (RF) chains. More recently, spatially movable antenna (SMA) architectures, including movable antenna (MA)~\cite{Zhu2024TWC,Zhu2025COMSTTutorialMA} and fluid antenna systems (FAS)~\cite{Wong2022TWCFAMA,New2025COMST}, have emerged to unlock additional spatial DoFs for MIMO communications. Although EMRA and SMA have each attracted much attention, their integration remains largely underexplored~\cite{Ma2026COMSTSurveyRMA}.

\subsection{Prior Work}

For EMRA architectures, one appealing benefit is the ability to adapt the radiation field~\cite{Hasan2018TWC,Bahceci2017TWC,Wang2023TWC}. By electronically tuning loads or switch states, the current distribution of each antenna element can be changed, thereby reshaping its radiation pattern. At the system level, this effect can be modeled through antenna-mode selection or basis-function expansions whose coefficients serve as electromagnetic (EM)-domain precoders~\cite{Zhao2021TWCMode,Wang2023TWC,Ying2024MWCRMMIMO,Ying2025TCOM}. Pattern-reconfigurable antennas have also been incorporated into tri-hybrid transceivers that jointly optimize digital, analog, and antenna-domain precoders under per-antenna power constraints~\cite{Zheng2026TCOMTriHybrid}. That work further treats both discrete and continuous pattern-reconfiguration models and validates the design with measured reconfigurable-antenna hardware data. In this setting, transmitter design depends not only on conventional spatial-domain channel state information (sCSI), but also on EM-domain channel state information (eCSI), which links the propagation geometry to the radiation-pattern basis and enables joint EM-domain and digital-domain precoding for substantial SE improvement over TFA systems~\cite{Wang2023TWC,Ying2025TCOM}.

One core challenge for EMRA systems is channel estimation (CE). Recent studies have also examined channel state information (CSI) extrapolation across radiation states~\cite{Liang2024TVTExtrap}, compressive CE for holographic RIS-aided terahertz massive MIMO~\cite{Wan2021TCOMTHzHRIS}, and hybrid knowledge-data driven channel acquisition and beamforming~\cite{Gao2023JSTSPSemantic}. In wideband parametric CE pipelines, the channel is first characterized by multipath delays, angle-of-departure (AoD), angle-of-arrival (AoA), and equivalent path gains, and the eCSI required for EM-domain design is then reconstructed from these geometric parameters~\cite{Bahceci2017TWC,Liao2019TCOM,Ying2025TCOM}. Although this representation reduces the estimation dimension, its accuracy still hinges on spatial sampling. In orthogonal frequency division multiplexing (OFDM) systems, frequency diversity mainly assists delay estimation, whereas reliable AoD recovery still requires sufficiently rich spatial observations and is further affected by spatial-wideband and frequency-wideband effects~\cite{Wang2018TSPWideband}. For fixed-geometry arrays, simply enlarging the inter-element spacing is not always effective. Although a larger aperture can improve angular resolution, excessive spacing may also introduce spatial aliasing and grating-lobe ambiguity~\cite{Liao2019TCOM}. Since AoD errors directly perturb the steering and basis matrices used to assemble eCSI, they propagate to EM-domain precoding and degrade the subsequent transmission design~\cite{Costa2010TSPSpherical,Nadeem2015TSP3D,Ying2025TCOM}.

Work on SMA architectures now spans channel modeling, antenna-position optimization, and multiuser precoding for MA and FAS systems~\cite{Zhu2024TWC,Yang2025TSPFWMMSE,Xiao2024TWCJointPositioning,Zhu2023LCOMMNullSteering}, and recent surveys provide broader accounts of these developments~\cite{Zhu2025COMSTTutorialMA,New2025COMST}. Antenna translation can also be combined with rotation to provide position and orientation DoFs~\cite{Ma2026COMSTSurveyRMA}. Linear minimum mean square error and Bayesian estimators have been studied for CE~\cite{Skouroumounis2023TCOM,Zhang2025TWC}. {Compressed sensing (CS) methods recover AoDs, AoAs, and path gains from measurements at selected MA positions~\cite{Ma2023LCOMMCSMA,Xiao2024TWCCSMA}; with an EM radiation basis, these parameters can be used to assemble eCSI. CS has also been applied to joint active-user detection and channel estimation in massive MIMO and to integrated sensing and communication~\cite{Ke2020TSPAUDCE,Gao2023TWCISACCS}. For the MA-specific methods, the angular estimates depend on the grid resolution and sensing geometry. In the proposed design, coordinated motion forms a uniform virtual array for two-dimensional estimation of signal parameters via rotational invariance techniques (ESPRIT), and the recovered parameters are used to assemble eCSI at the transmission positions.}

Recent studies have examined the integration of spatial and EM reconfiguration through unified system models and signal processing methods supported by artificial intelligence~\cite{Ma2026COMSTSurveyRMA,Li2026ArxivAIMARA}. Pixel-based reconfigurable antennas emulate FAS ports on a fixed aperture~\cite{Zhang2025OJAPPRAFAS}, and the reconfigurable pixel antenna (RPA)-based electronic movable-antenna array (REMAA) selects from a finite set of candidate radiation positions for multiuser transmission~\cite{Chen2025TCOMREMAA}. A radiation-center reconfigurable antenna array (RCRAA) similarly combines discrete radiation-center selection with analog and digital beamforming~\cite{Li2026TWCRCRAA}. An electromagnetically reconfigurable FAS further combines software-controlled fluidics with radiation-pattern control and has been validated through full-wave simulations, fabricated prototypes, and communication experiments~\cite{Wang2026JSACERFAS}. Together, these architectures provide spatial and EM flexibility through effective positions, radiation centers, fluidic control, and radiation-pattern reconfiguration~\cite{Li2026JSACRPRFAS}.

{Physical repositioning directly controls the sampling coordinates and supplies a regular observation geometry for virtual-array CE. The EM coefficients remain available for radiation-pattern adaptation during data transmission, so sensing and transmission retain separate control variables~\cite{Zhang2025OJAPPRAFAS,Chen2025TCOMREMAA,Li2026TWCRCRAA}.}

{The spatial-EM reconfigurable antenna (SEMRA) architecture considered here coordinates antenna repositioning across pilot slots to form a denser virtual array under the same per-user pilot overhead as EMRA. The resulting parametric eCSI model supports tri-domain precoding over antenna positions, EM-domain pattern weights, and digital precoders.}

{Table~\ref{tab:architecture_comparison} compares representative MIMO architectures in terms of digital precoding, analog precoding, EM precoding, and spatial reposition.}

\begin{table*}[!t]
	
	\centering
	\caption{Comparison of reconfiguration domains and mechanisms in representative MIMO architectures.}
	\label{tab:architecture_comparison}
	\renewcommand{\arraystretch}{1.22}
	\setlength{\tabcolsep}{4pt}
	\footnotesize
	\begin{tabularx}{\textwidth}{@{}
		>{\raggedright\arraybackslash}p{0.132\textwidth}
		>{\centering\arraybackslash}p{0.046\textwidth}
		>{\centering\arraybackslash}p{0.078\textwidth}
		>{\centering\arraybackslash}p{0.078\textwidth}
		>{\centering\arraybackslash}p{0.078\textwidth}
		>{\centering\arraybackslash}p{0.078\textwidth}
		>{\centering\arraybackslash}X
		>{\centering\arraybackslash}X
		@{}}
		\toprule
		\textbf{Architecture} &
		\textbf{Ref.} &
		\textbf{\shortstack{Digital\\precoding}} &
		\textbf{\shortstack{Analog\\precoding}} &
		\textbf{\shortstack{EM\\precoding}} &
		\textbf{\shortstack{Spatial\\reposition}} &
		\textbf{\shortstack{EM reconfigurable\\mechanism}} &
		\textbf{\shortstack{Spatial reconfigurable\\mechanism}} \\
		\midrule
		Hybrid MIMO &
		\cite{Gao2023TWCISACCS} &
		\(\checkmark\) &
		\(\checkmark\) & & & & \\
		\midrule
		\multirow{2}{0.132\textwidth}{EM-Reconfigurable MIMO} &
		\cite{Ying2025TCOM} &
		\(\checkmark\) &
		&
		\(\checkmark\) & &
		SH weighting & \\
		\cmidrule(l){2-8}
		&
		\cite{Li2026JSACRPRFAS} &
		\(\checkmark\) &
		&
		\(\checkmark\) & &
		BD/BW tuning & \\
		\midrule
		\multirow{2}{0.132\textwidth}{Movable/fluid antenna} &
		\cite{Wong2022TWCFAMA} &
		& & &
		\(\checkmark\) & &
		Preset port selection \\
		\cmidrule(l){2-8}
		&
		\cite{Zhu2024TWC} &
		\(\checkmark\) & & &
		\(\checkmark\) & &
		Continuous position optimization \\
		\midrule
		\multirow{3}{0.132\textwidth}{Tri-hybrid MIMO} &
		\cite{Ying2024MWCRMMIMO} &
		\(\checkmark\) &
		\(\checkmark\) &
		\(\checkmark\) & &
		Pixel pattern switching & \\
		\cmidrule(l){2-8}
		&
		\cite{Li2026TWCRCRAA} &
		\(\checkmark\) &
		\(\checkmark\) &
		\(\checkmark\) & &
		Discrete RC selection & \\
		\cmidrule(l){2-8}
		&
		\cite{Zheng2026TCOMTriHybrid} &
		\(\checkmark\) &
		\(\checkmark\) &
		\(\checkmark\) & &
		General pattern selection/synthesis & \\
		\midrule
		Proposed tri-domain MIMO &
		Our work &
		\(\checkmark\) &
		&
		\(\checkmark\) &
		\(\checkmark\) &
		SH weighting &
		Continuous position optimization \\
		\bottomrule
	\end{tabularx}
	\par\addvspace{5mm}
	\begin{minipage}{\textwidth}
	\footnotesize
	Note. In the mechanism columns, SH denotes spherical-harmonic, BD denotes beam direction, BW denotes beamwidth, and RC denotes radiation center.
	\end{minipage}
\end{table*}

\subsection{Contributions}

This paper presents a SEMRA framework that combines movement-aided parametric CE with tri-domain precoding. Antenna repositioning supplies the spatial samples for virtual-array eCSI reconstruction and the position variables for transmission optimization, while EM-domain pattern weights and digital precoders are optimized for data transmission. The contributions are summarized as follows.

\begin{itemize}
\item Within the proposed SEMRA framework, we develop a concrete low-overhead movement-aided parametric CE procedure. By coordinating local repositioning of the $M_y\times M_z$ antennas over a prescribed multi-block pilot schedule, a denser virtual array is synthesized while preserving the same per-user pilot overhead as the EMRA baseline. The resulting virtual array supplies additional spatial samples for the parametric estimation pipeline, improves AoD estimation, and yields a position-independent parametric channel model from which the eCSI can be assembled at desired antenna positions without additional pilots.

\item We formulate the SEMRA SE-maximization problem and develop a tri-domain alternating optimization framework based on block coordinate descent (BCD) over antenna positions, EM-domain pattern weights, and digital precoders. A zero-forcing (ZF)-based joint design is first constructed as a low-complexity baseline to isolate the gain of spatial reconfiguration beyond existing EM-digital designs for EMRA systems. We then develop a weighted minimum mean square error (WMMSE)-based design to further improve the system SE.

\item Numerical results show that the proposed CE scheme improves eCSI accuracy over the EMRA baseline under the same per-user pilot overhead. Within the tested range, both SEMRA-ZF and SEMRA-WMMSE achieve higher SE than the EMRA baseline under both estimated and perfect eCSI scenarios. For the tested multiuser loads, SEMRA-WMMSE further improves over SEMRA-ZF. The SEMRA-over-EMRA advantage is especially evident at larger inter-element spacings, where the EMRA baseline is more sensitive to spatial aliasing.
\end{itemize}

{The remainder of this paper is organized as follows. Section~\ref{S2} presents the system model and formulates the tri-domain SE-maximization problem. Sections~\ref{S3} and~\ref{S4} develop the SEMRA-ZF and SEMRA-WMMSE designs, respectively. Section~\ref{S5} introduces the movement-aided parametric CE procedure. Section~\ref{S6} reports the simulation results, and Section~\ref{S7} concludes the paper.}

\textit{Notation}: Matrices and column vectors are denoted by uppercase and lowercase boldface letters, respectively. $(\cdot)^{T}$, $(\cdot)^{*}$, $(\cdot)^{H}$, $(\cdot)^{-1}$, $(\cdot)^{\dagger}$, and $\mathbb{E}\{\cdot\}$ denote the transpose, conjugate, Hermitian transpose, inversion, Moore-Penrose pseudoinverse, and expectation. $[\bm{x}]_k$ and $[\bm{A}]_{i,j}$ denote the $k$-th entry of vector $\bm{x}$ and the $(i,j)$-th entry of matrix $\bm{A}$. $\mathrm{diag}(\bm{a})$ forms a diagonal matrix from vector $\bm{a}$, and $\mathrm{Blkdiag}\{\bm{a}_1,\ldots,\bm{a}_M\}$ constructs a block diagonal matrix. $\odot$ denotes the Hadamard product. $\|\cdot\|_2$ and $\|\cdot\|_{\mathrm{F}}$ denote the $\ell_2$ norm and Frobenius norm, respectively. $\Re\{\cdot\}$ and $\Im\{\cdot\}$ denote the real and imaginary parts, respectively. $\mathrm{j}$ is the imaginary unit. $\bm{I}_N$ denotes the $N \times N$ identity matrix. $\mathcal{CN}(\mu,\sigma^2)$ denotes a complex Gaussian distribution.

\section{System Model and Problem Formulation}\label{S2}
\subsection{Downlink Signal Model}\label{S2.1}
\begin{figure}[!t]
	\centering
	\includegraphics[width=\columnwidth]{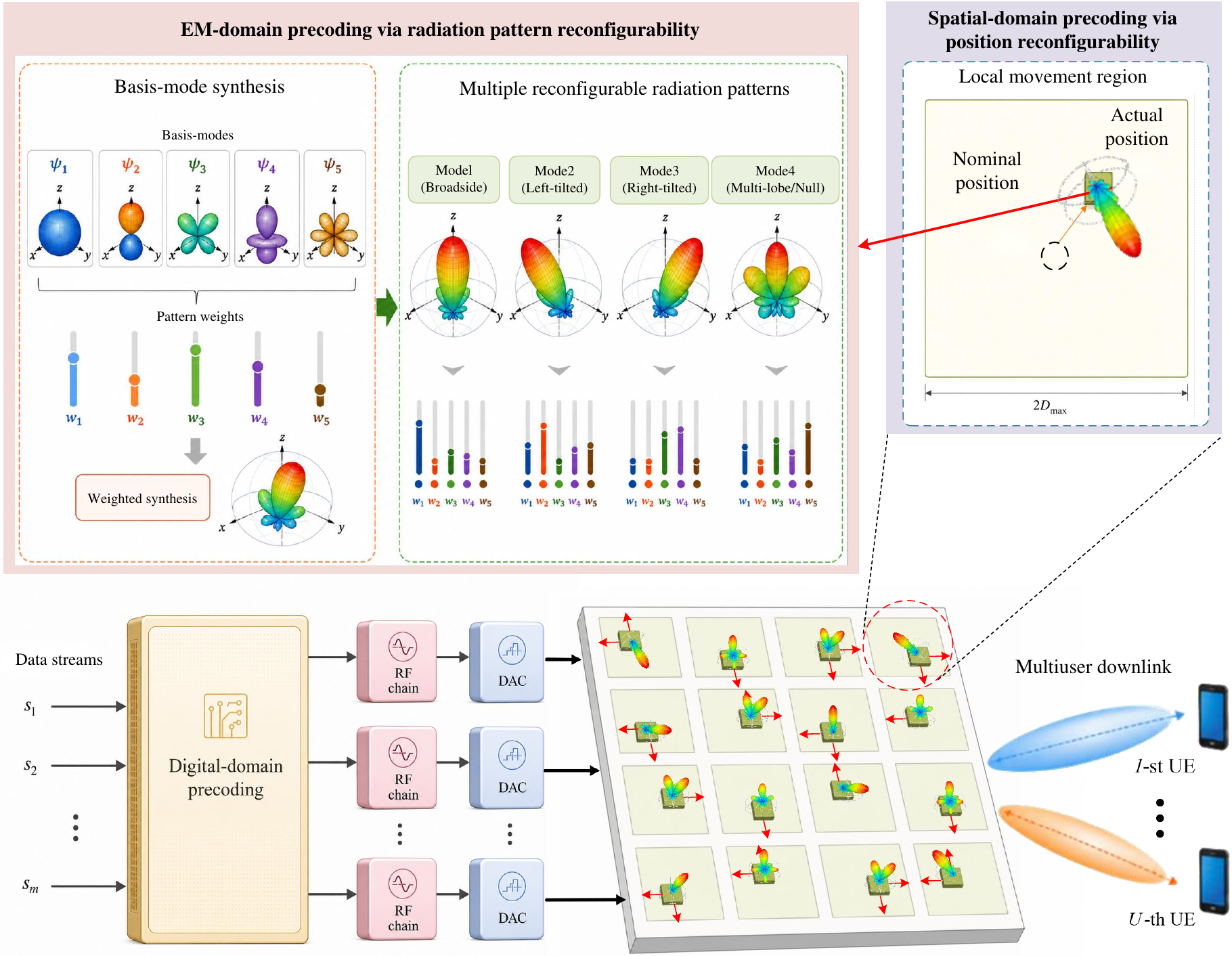}
	\caption{SEMRA-enabled multiuser downlink with EM-domain radiation-pattern reconfigurability and spatial-domain position reconfigurability.}
	\label{fig:system_model}
	\vspace*{-3mm}
\end{figure}
We consider a time-division duplex (TDD) multiuser MIMO-OFDM downlink system implemented with the SEMRA architecture illustrated in Fig.~\ref{fig:system_model}. The base station (BS) employs an $M=M_yM_z$ uniform planar array (UPA) of SEMRA elements to serve $U$ single-antenna user equipments (UEs), each with a fixed position and radiation pattern. The inter-element spacings of the UPA along the y- and z-axes are $d_y$ and $d_z$, respectively. The number of OFDM subcarriers is $G$ with subcarrier spacing $\Delta f \triangleq B_w/G$, and $f_g \triangleq (g-1)\Delta f$ ($1 \le g \le G$) denotes the frequency of the $g$-th subcarrier. The received signal at the $u$-th UE ($1 \le u \le U$) and the $g$-th subcarrier ($1 \le g \le G$) is given by
\begin{equation}\label{equ.rx}
    y_{u,g} = \bm{h}_{u,g}^{H} \bm{W}_g \bm{s}_g + n_{u,g},
\end{equation}
where $\bm{h}_{u,g} = [h_{u,1,g}, \ldots, h_{u,M,g}]^T \in \mathbb{C}^{M}$ is the downlink channel vector between the BS and the $u$-th UE, $\bm{W}_g = [\bm{w}_{1,g}, \ldots, \bm{w}_{U,g}] \in \mathbb{C}^{M \times U}$ is the digital precoding matrix with $\bm{w}_{u,g} \in \mathbb{C}^{M \times 1}$ denoting the digital precoder vector for the $u$-th UE on the $g$-th subcarrier, $\bm{s}_g \in \mathbb{C}^{U}$ is the data symbol vector satisfying $\mathbb{E}\{\bm{s}_g \bm{s}_g^{H}\} = \bm{I}_U$, and $n_{u,g} \sim \mathcal{CN}(0, \sigma_n^2)$ is additive white Gaussian noise (AWGN).

\subsection{Channel Model}\label{S2.2}
The channel between the $m$-th BS antenna and the $u$-th UE on the $g$-th subcarrier is modeled as~\cite{38.901}
\begin{align}
\begin{split}
\label{equ.ch}
    h_{u,m,g} &= \sum_{i=1}^{L_u} \tilde{x}_{i,u} f_{\mathrm{rx},u}(\vartheta_{i,u}, \varphi_{i,u}) f_{\mathrm{tx},m}(\theta_{i,u}, \phi_{i,u}) \\
    &\times e^{-\mathrm{j}\frac{2\pi}{\lambda}(\bm{k}_{\mathrm{tx},i,u}^T \bm{p}_m + \bm{k}_{\mathrm{rx},i,u}^T \bm{r}_u)} \cdot e^{-\mathrm{j}2\pi \tau_{i,u} f_g},
\end{split}
\end{align}
where $L_u$ is the number of channel paths, $\tilde{x}_{i,u}$ and $\tau_{i,u}$ denote the complex gain and delay of the $i$-th path, respectively, $(\theta_{i,u}, \phi_{i,u})$ and $(\vartheta_{i,u}, \varphi_{i,u})$ are respectively the AoD and AoA, $f_{\mathrm{tx},m}(\cdot)$ and $f_{\mathrm{rx},u}(\cdot)$ denote the real-valued scalar radiation-pattern gains of the BS and UE antennas, respectively, and $\bm{p}_m$, $\bm{r}_u$ are the position vectors of BS antenna~$m$ and UE~$u$, respectively. Throughout this paper, the BS-side departure azimuth is restricted to the front half-space, i.e., $\phi_{i,u}\in[-\pi/2,\pi/2]$. The unit direction vectors $\bm{k}_{\mathrm{tx},i,u}, \bm{k}_{\mathrm{rx},i,u} \in \mathbb{R}^3$ are defined as $\bm{k}_{\mathrm{tx},i,u} = [\sin\theta_{i,u}\cos\phi_{i,u}, \sin\theta_{i,u}\sin\phi_{i,u}, \cos\theta_{i,u}]^T$ and $\bm{k}_{\mathrm{rx},i,u} = [\sin\vartheta_{i,u}\cos\varphi_{i,u}, \sin\vartheta_{i,u}\sin\varphi_{i,u}, \cos\vartheta_{i,u}]^T$.

Incorporating the delay into the path gain as $x_{i,u,g} = \tilde{x}_{i,u} e^{-\mathrm{j}2\pi\tau_{i,u}f_g}$, the channel can be written in the compact matrix form as
\begin{equation}\label{equ.ch_compact}
    h_{u,m,g} = \bm{f}_{\mathrm{rx},u}^T \bm{A}_u \bm{\Sigma}_{u,g} \bm{B}_{u,m} \bm{f}_{\mathrm{tx},u,m},
\end{equation}
where $\bm{f}_{\mathrm{rx},u} \in \mathbb{R}^{L_u}$ and $\bm{f}_{\mathrm{tx},u,m} \in \mathbb{R}^{L_u}$ collect the receive and transmit pattern gains along all $L_u$ paths, with the $i$-th elements being $f_{\mathrm{rx},u}(\vartheta_{i,u},\varphi_{i,u})$ and $f_{\mathrm{tx},m}(\theta_{i,u},\phi_{i,u})$, respectively, $\bm{A}_u = \mathrm{diag}(\bm{a}_u) \in \mathbb{C}^{L_u \times L_u}$ and $\bm{B}_{u,m} = \mathrm{diag}(\bm{b}_{u,m}) \in \mathbb{C}^{L_u \times L_u}$ capture the array steering phases, with $[\bm{a}_u]_i = e^{-\mathrm{j}\frac{2\pi}{\lambda}\bm{k}_{\mathrm{rx},i,u}^T \bm{r}_u}$ and $[\bm{b}_{u,m}]_i = e^{-\mathrm{j}\frac{2\pi}{\lambda}\bm{k}_{\mathrm{tx},i,u}^T \bm{p}_m}$, and $\bm{\Sigma}_{u,g} = \mathrm{diag}([x_{1,u,g}, \ldots, x_{L_u,u,g}]) \in \mathbb{C}^{L_u \times L_u}$ collects the per-path gains.

\subsection{EM-Domain CSI Representation}\label{S2.3}
To enable EM-domain radiation pattern optimization, we represent the transmit pattern as a weighted combination of $K$ orthonormal basis functions
\begin{equation}\label{equ.sh}
    f_{\mathrm{tx},m}(\theta, \phi) = \sum_{k=1}^{K} \alpha_{k,m} \omega_k(\theta, \phi),
\end{equation}
where $\{\omega_k(\theta, \phi)\}_{k=1}^{K}$ are real-valued orthonormal basis functions over the unit sphere, constructed from real spherical harmonics as detailed in~\cite{Ying2025TCOM}. The orthonormality condition reads $\int_{0}^{2\pi}\!\int_{0}^{\pi}\omega_k(\theta,\phi)\omega_{k'}(\theta,\phi)\sin\theta\,\mathrm{d}\theta\,\mathrm{d}\phi = \delta_{k,k'}$,
and $\bm{\alpha}_m = [\alpha_{1,m}, \ldots, \alpha_{K,m}]^T \in \mathbb{R}^K$ is the weight coefficient vector for the $m$-th antenna. Under the adopted decomposition, $f_{\mathrm{tx},m}(\theta,\phi)$ models the real-valued directional gain, and we further assume that $\|\bm{\alpha}_m\|_2 = 1$. Applying \eqref{equ.sh} to the transmit pattern gain vector yields
\begin{equation}\label{equ.ftx}
    \bm{f}_{\mathrm{tx},u,m} = \bm{\Omega}_u \bm{\alpha}_m \in {\mathbb{R}^{L_u}}, \quad \forall u, m,
\end{equation}
where $\bm{\alpha}_m$ parameterizes real gain-pattern coefficients, and the $(i,k)$-th element of $\bm{\Omega}_u \in \mathbb{R}^{L_u \times K}$ is $[\bm{\Omega}_u]_{i,k} = \omega_k(\theta_{i,u}, \phi_{i,u})$. Substituting \eqref{equ.ftx} into \eqref{equ.ch_compact}, the channel can be expressed as
\begin{equation}\label{equ.ecsi}
    h_{u,m,g} = \bm{f}_{\mathrm{rx},u}^T \bm{A}_u \bm{\Sigma}_{u,g} \bm{B}_{u,m} \bm{\Omega}_u \bm{\alpha}_m
    = \bm{q}_{u,m,g}^H \bm{\alpha}_m.
\end{equation}
We define the eCSI vector as
\begin{equation}\label{equ.q_def}
    \bm{q}_{u,m,g}^H \triangleq \bm{f}_{\mathrm{rx},u}^T \bm{A}_u \bm{\Sigma}_{u,g} \bm{B}_{u,m} \bm{\Omega}_u \in \mathbb{C}^{1 \times K},
\end{equation}
and define the corresponding position-independent equivalent channel gain vector as
\begin{equation}\label{equ.v_def}
    \bm{\chi}_{u,g} \triangleq \left(\bm{f}_{\mathrm{rx},u}^T \bm{A}_u \bm{\Sigma}_{u,g}\right)^H \in \mathbb{C}^{L_u}.
\end{equation}
It collects the position-independent UE-side and path-gain terms. The corresponding column-vector form of the eCSI is
\(
\bm{q}_{u,m,g} = \bm{\Omega}_u^T \bm{B}_{u,m}^H \bm{\chi}_{u,g}
\in \mathbb{C}^{K\times 1}.
\)

By aggregating the eCSI across all $M$ antennas, we define the stacked eCSI vector $\bm{q}_{u,g} \triangleq [\bm{q}_{u,1,g}^T, \ldots, \bm{q}_{u,M,g}^T]^T \in \mathbb{C}^{MK \times 1}$ and its conjugated form $\bar{\bm{q}}_{u,g} \triangleq \bm{q}_{u,g}^*$. We also define the block-diagonal EM precoding matrix $\bm{\Lambda} \triangleq \mathrm{Blkdiag}\{\bm{\alpha}_1, \ldots, \bm{\alpha}_M\} \in \mathbb{R}^{MK \times M}$. The effective channel then satisfies
\begin{equation}\label{equ.h_def}
    \bm{h}_{u,g}^H = \bar{\bm{q}}_{u,g}^H \bm{\Lambda}.
\end{equation}
Substituting~\eqref{equ.h_def} into~\eqref{equ.rx} yields $y_{u,g} = \bar{\bm{q}}_{u,g}^H \bm{\Lambda} \bm{W}_g \bm{s}_g + n_{u,g}$, so the signal-to-interference-plus-noise ratio (SINR) for the $u$-th UE on the $g$-th subcarrier is
\begin{equation}\label{equ.sinr}
    \mathrm{SINR}_{u,g} = \frac{|\bar{\bm{q}}_{u,g}^H \bm{\Lambda} \bm{w}_{u,g}|^2}{\sum_{\substack{\ell=1 \\ \ell \neq u}}^{U} |\bar{\bm{q}}_{u,g}^H \bm{\Lambda} \bm{w}_{\ell,g}|^2 + \sigma_n^2}.
\end{equation}
Accordingly, the sum SE is given by
\begin{equation}\label{equ.se}
    R = \sum_{g=1}^{G} \sum_{u=1}^{U} \log_2 (1 + \mathrm{SINR}_{u,g}).
\end{equation}

\subsection{SEMRA Model and Problem Formulation}\label{S2.4}
Unlike the EMRA with TFA architecture, the proposed SEMRA architecture jointly optimizes the per-antenna EM-domain pattern weights $\bm{\alpha}_m$ and antenna positions $\bm{p}_m$. Let $\bar{\bm{p}}_m$ denote the reference position of the $m$-th antenna in the reference UPA. In SEMRA, each transmit antenna is permitted to move within a local feasible region $\mathcal{S}_m$ centered at $\bar{\bm{p}}_m$ (depicted as the local movement region in Fig.~\ref{fig:system_model}), which is defined as
\begin{equation}\label{equ.constraint}
    \bm{p}_m \in \mathcal{S}_m \triangleq \left\{ \bm{x} \in \mathbb{R}^3 \mid \|\bm{x} - \bar{\bm{p}}_m\|_\infty \le D_{\max},\ [\bm{x}]_1 = 0 \right\},
\end{equation}
where $D_{\max}$ is the maximum allowable displacement for each antenna. To avoid antenna collisions, we require a minimum inter-antenna separation $D_{\mathrm{sep}} > 0$ over all feasible positions, which is ensured for the reference UPA if $d_y \ge 2D_{\max} + D_{\mathrm{sep}}$ and $d_z \ge 2D_{\max} + D_{\mathrm{sep}}$. The position $\bm{p}_m$ enters the channel through the BS-side steering matrix $\bm{B}_{u,m}$ in~\eqref{equ.ch_compact}, introducing an additional spatial degree of freedom for performance optimization.

According to~\eqref{equ.h_def}, the effective channels are jointly determined by the antenna positions $\bm{p}\triangleq[\bm{p}_1^T,\ldots,\bm{p}_M^T]^T \in \mathbb{R}^{3M}$ and the per-antenna EM-domain weights $\bm{\alpha}_m$, whose effects are captured by $\bar{\bm{q}}_{u,g}$ and $\bm{\Lambda}$, respectively. Let $\bm{W}\triangleq\{\bm{W}_g\}_{g=1}^{G}$ denote the collection of digital precoders across all subcarriers. We aim to maximize the sum SE of all UEs in~\eqref{equ.se} by jointly optimizing digital-domain precoder $\bm{W}$, EM-domain radiation-pattern weights $\bm{\alpha}_m$, and spatial-domain antenna position $\bm{p}$, leading to
\begin{align}\label{equ.problem}
\begin{aligned}
    \mathcal{P}_{\mathrm{SE}}: \quad \max_{\{\bm{\alpha}_m\}, \bm{W}, \bm{p}} \quad & R \\
    \text{s.t.} \quad & \sum_{g=1}^{G} \|\bm{W}_g\|_{\mathrm{F}}^2 \le P_T, \\
    & \|\bm{\alpha}_m\|_2 = 1,\ \bm{p}_m \in \mathcal{S}_m,\ \forall m.
\end{aligned}
\end{align}
where $R$ is defined in \eqref{equ.se} and depends on $\bm{W}$, $\bm{\alpha}_m$, and $\bm{p}$ through the effective channel vectors $\bm{h}_{u,g}$ in \eqref{equ.h_def}. The resulting problem is highly nonconvex due to the coupled variables and the unit-norm constraints, making it challenging to solve optimally.

\begin{figure*}[!t]
	\centering
	\includegraphics[width=0.95\textwidth]{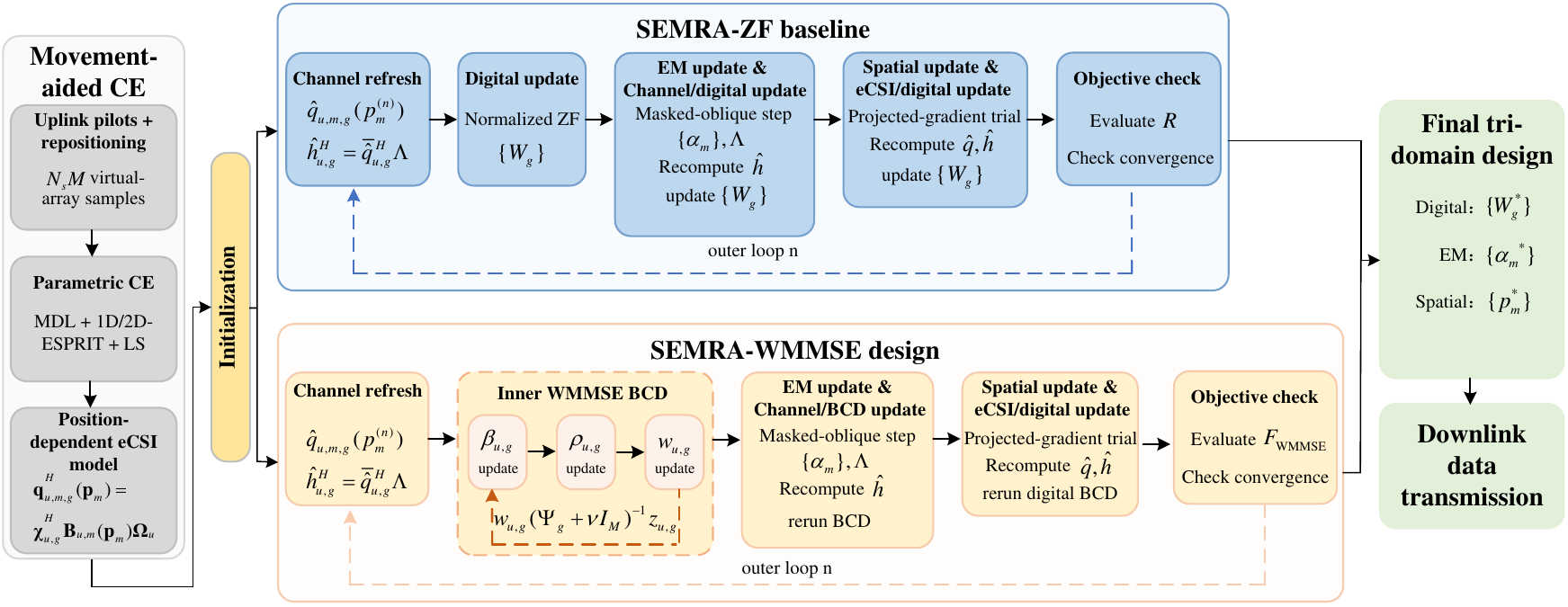}
	\caption{Concrete workflow of the proposed SEMRA method, including movement-aided CE, position-dependent eCSI reconstruction, and ZF/WMMSE tri-domain precoding.}
	\label{fig:semra_workflow}
	\vspace*{-3mm}
\end{figure*}

\section{Tri-Domain Alternating Optimization: ZF-Based Baseline}\label{S3}

Building upon the EM-digital alternating ZF design in~\cite{Ying2025TCOM}, we extend it to SEMRA by adding a spatial-domain position-reconfigurable block. The detailed CE-to-transmission workflow is summarized in Fig.~\ref{fig:semra_workflow}, with the ZF, WMMSE, and CE components developed in Sections~\ref{S3}-\ref{S5}, respectively. 

{SEMRA-ZF is used as the lower complexity reference design. Its normalized ZF update suppresses multiuser interference and makes the gain of the added spatial block directly comparable with the EM and digital baseline, but it does not optimize user power allocation under the global power constraint.}

\subsection{Digital and EM-Domain Optimization}\label{S3.1zf}

Given the current effective channels $\bm{h}_{u,g}$ in~\eqref{equ.h_def}, we adopt the ZF precoder. Define the normalized channel direction and the stacked matrix as in~\eqref{equ.O_g_def}.
\begin{equation}\label{equ.O_g_def}
    \tilde{\bm{h}}_{u,g} \triangleq \frac{\bm{h}_{u,g}}{\|\bm{h}_{u,g}\|_2}, \qquad
    \tilde{\bm{H}}_g \triangleq [\tilde{\bm{h}}_{1,g},\ldots,\tilde{\bm{h}}_{U,g}] \in \mathbb{C}^{M\times U}.
\end{equation}
For $M \ge U$, the normalized-channel ZF precoder is computed as 
\begin{equation}\label{equ.zf_normalize}
    \bm{W}_g = \sqrt{\frac{P_T}{G}} \frac{\tilde{\bm{W}}_g}{\|\tilde{\bm{W}}_g\|_{\mathrm{F}}},
\end{equation}
where $\tilde{\bm{W}}_g \triangleq \tilde{\bm{H}}_g \big(\tilde{\bm{H}}_g^H \tilde{\bm{H}}_g\big)^{-1} \in \mathbb{C}^{M\times U}$ is the unnormalized ZF precoder built from the normalized channel directions in~\eqref{equ.O_g_def}.

With $\bm{p}$ and the digital precoders fixed, we update the EM-domain precoder by optimizing the block-diagonal matrix $\bm{\Lambda}=\mathrm{Blkdiag}\{\bm{\alpha}_1,\ldots,\bm{\alpha}_M\}$ under per-antenna unit-norm constraints. This is formulated on a masked oblique manifold. We define the oblique manifold
\begin{equation}\label{equ.oblique_def}
    \mathcal{OB} \triangleq \left\{\bm{\Lambda}\in \mathbb{R}^{KM\times M}: \left[\bm{\Lambda}^{T}\bm{\Lambda}\right]_{m,m} = 1, \; \forall m \right\}.
\end{equation}
Let $\bm{M}_{0} \triangleq \mathrm{Blkdiag}\{\bm{1}_{K},\ldots, \bm{1}_{K}\}\in\mathbb{R}^{KM\times M}$ and impose the structural constraint $\bm{\Lambda}=\bm{\Lambda}\odot \bm{M}_{0}$.
Equivalently, the feasible EM set is $\{\bm{\Lambda}\in\mathcal{OB}:\bm{\Lambda}=\bm{\Lambda}\odot\bm{M}_{0}\}$.
We define the cost function $\mathfrak{f}(\bm{\Lambda})\triangleq -R$. The Riemannian gradient at $\bm{\Lambda}$ is the orthogonal projection of the Euclidean gradient $\nabla \mathfrak{f}(\bm{\Lambda})$ onto the tangent space, given by
\begin{equation}\label{equ.oblique_grad}
    \mathrm{grad}\mathfrak{f}(\bm{\Lambda}) = \nabla \mathfrak{f}(\bm{\Lambda}) - \bm{\Lambda}\,\mathrm{ddiag}\!\left(\bm{\Lambda}^{T}\nabla \mathfrak{f}(\bm{\Lambda})\right),
\end{equation}
where $\mathrm{ddiag}(\cdot)$ sets all off-diagonal entries of its matrix argument to zero. The retraction is column-wise normalization $\mathrm{Retr}_{\bm{\Lambda}}(\gamma \bm{D}) \triangleq \mathrm{normalize}(\bm{\Lambda}+\gamma \bm{D})$.

We adopt the Riemannian steepest-descent direction on this masked-oblique set, namely
\begin{equation}\label{equ.oblique_cg_dir}
    \bm{D}^{(n)} = -\mathrm{grad}\mathfrak{f}\!\left(\bm{\Lambda}^{(n)}\right),
\end{equation}
and choose a suitable step size $\gamma^{(n)}$ for the current EM subproblem with the digital precoders fixed. The EM precoder update is then
\begin{equation}\label{equ.lambda_update_zf}
\begin{split}
    \bm{\Lambda}^{(n+1)}
    &= \mathrm{Retr}_{\bm{\Lambda}^{(n)}}\!\left(\gamma^{(n)} \bm{D}^{(n)}\right) \\
    &= \mathrm{normalize}\!\left(\bm{\Lambda}^{(n)}+\gamma^{(n)}\bm{D}^{(n)}\right).
\end{split}
\end{equation}
To keep the per-outer-iteration cost moderate, the EM block in both tri-domain algorithms below uses a single manifold-gradient step in each outer iteration rather than an inner loop to full convergence. Since the digital and spatial blocks are refreshed between successive EM updates, the EM descent direction is recomputed from the current gradient in every outer iteration. For compactness, variables without outer-iteration superscripts in the algorithm below denote the latest available block values within the current outer iteration.
To express the Euclidean gradient compactly, we use~\eqref{equ.h_def}, i.e., $\bm{h}_{u,g}^H = \bar{\bm{q}}_{u,g}^H \bm{\Lambda}$. Let $\zeta \triangleq 1/\sigma_n^2$, and $\bm{W}_{\bar{u},g}$ denote the digital precoder matrix obtained from $\bm{W}_g$ by removing its $u$-th column. The Euclidean gradient $\nabla \mathfrak{f}(\bm{\Lambda})$ is given by~\cite{Ying2025TCOM}
\begin{equation}\label{equ.grad_euc_zf_lambda}
    \nabla \mathfrak{f}\!\left(\bm{\Lambda}\right)
    = \left( -\sum_{g=1}^{G}\sum_{u=1}^{U}\frac{ \bm{\Gamma}_{u,g}^{(1)} -\bm{\Gamma}_{u,g}^{(2)}}{\ln 2} \right) \odot \bm{M}_{0}.
\end{equation}
The Hadamard masking enforces the block-diagonal sparsity of $\bm{\Lambda}$, such that each column update only acts on its corresponding $K$-dimensional pattern-weight block. Starting from a feasible block-diagonal initialization, the masked gradient in~\eqref{equ.grad_euc_zf_lambda} and the retraction in~\eqref{equ.lambda_update_zf} preserve this structure throughout the ZF-EM updates. Moreover,
\begin{align}\label{equ.Gamma_zf_def}
    \bm{\Gamma}_{u,g}^{(1)} &\triangleq \frac{2\zeta \Re\!\left\{\bar{\bm{q}}_{u,g}\bar{\bm{q}}_{u,g}^{H}\bm{\Lambda}\bm{W}_g\bm{W}_g^{H}\right\}}{1+\zeta\|\bar{\bm{q}}_{u,g}^{H}\bm{\Lambda}\bm{W}_{g}\|_2^{2}}, \\
    \bm{\Gamma}_{u,g}^{(2)} &\triangleq \frac{2\zeta \Re\!\left\{\bar{\bm{q}}_{u,g}\bar{\bm{q}}_{u,g}^{H}\bm{\Lambda}\bm{W}_{\bar{u},g}\bm{W}_{\bar{u},g}^{H}\right\}}{1+\zeta\|\bar{\bm{q}}_{u,g}^{H}\bm{\Lambda}\bm{W}_{\bar{u},g}\|_2^{2}}.
\end{align}

\subsection{Spatial-Domain Antenna Position Optimization}\label{S3.2zf}

Using the available parametric eCSI model, we derive the spatial gradient of the SE objective $R$ with respect to the \emph{conjugate} channel coefficient $h_{u,m,g}^*$.

\subsubsection{Multipath Channel Decomposition}
Based on~\eqref{equ.v_def} and $h_{u,m,g}^* = \bm{\alpha}_m^T \bm{q}_{u,m,g}$, the conjugate channel admits the pathwise decomposition
\begin{equation}\label{equ.h_decomp_zf}
    h_{u,m,g}^* = \bm{\alpha}_m^T \bm{\Omega}_u^T \bm{B}_{u,m}^H \bm{\chi}_{u,g}
    = \sum_{i=1}^{L_u} h_{u,m,g,i}^*,
\end{equation}
where the $i$-th path component is
\begin{equation}\label{equ.v_def_zf}
    h_{u,m,g,i}^* \triangleq [\bm{\Omega}_u \bm{\alpha}_m]_i\,[\bm{\chi}_{u,g}]_i\, e^{\mathrm{j}\frac{2\pi}{\lambda}\bm{k}_{\mathrm{tx},i,u}^T \bm{p}_m}.
\end{equation}
Here $\bm{k}_{\mathrm{tx},i,u}$ is determined by the AoD pair $(\theta_{i,u},\phi_{i,u})$, while $\bm{\chi}_{u,g}$ absorbs the remaining position-independent factors. Therefore, the spatial update depends only on the position-dependent eCSI model, not on explicit AoA or UE-side path parameters.

\subsubsection{Sum-SE Gradient Kernel}
Define the effective coupling coefficient from the $\ell$-th column of $\bm{W}_g$ to the $u$-th UE on the g-th subcarrier as
\begin{equation}\label{equ.c_def}
    c_{u,\ell,g} \triangleq \bm{h}_{u,g}^H \bm{w}_{\ell,g}, \quad \ell\in\{1,\ldots,U\}.
\end{equation}
In particular, $c_{u,u,g}$ corresponds to the desired signal term of the $u$-th UE, whereas $c_{u,\ell,g}$ for $\ell\neq u$ represent inter-user interference. Let $P_{u,g} \triangleq \sum_{\ell=1}^{U} |c_{u,\ell,g}|^2 + \sigma_n^2$ and $\mathcal{I}_{u,g} \triangleq \sum_{\substack{\ell=1 \\ \ell \neq u}}^{U} |c_{u,\ell,g}|^2 + \sigma_n^2$, so that $\mathrm{SINR}_{u,g}=|c_{u,u,g}|^2/\mathcal{I}_{u,g}$, $1+\mathrm{SINR}_{u,g} = P_{u,g}/\mathcal{I}_{u,g}$ and $r_{u,g} = \log_2(P_{u,g}) - \log_2(\mathcal{I}_{u,g})$. Moreover, $\frac{\partial |c_{u,\ell,g}|^2}{\partial h_{u,m,g}^*} = c_{u,\ell,g}^* w_{\ell,m,g}$ with $w_{\ell,m,g} = [\bm{w}_{\ell,g}]_m$. Applying the chain rule yields the sensitivity of $R$ with respect to $h_{u,m,g}^*$ as
\begin{align}\label{equ.xi_zf_def}
    \xi_{u,m,g}^{(R)} &\triangleq \frac{\partial R}{\partial h_{u,m,g}^*} = \frac{1}{\ln 2 \cdot P_{u,g}} \bigg( c_{u,u,g}^* w_{u,m,g} \nonumber\\
    &\quad - \mathrm{SINR}_{u,g} \textstyle\sum_{\substack{\ell=1 \\ \ell \neq u}}^{U} c_{u,\ell,g}^* w_{\ell,m,g} \bigg).
\end{align}
\subsubsection{Spatial Gradient Assembly and Projection}
From~\eqref{equ.v_def_zf}, differentiating the conjugate path component $h_{u,m,g,i}^*$ with respect to $\bm{p}_m$ yields $\frac{\partial h_{u,m,g,i}^*}{\partial \bm{p}_m} = \mathrm{j}\frac{2\pi}{\lambda} \bm{k}_{\mathrm{tx},i,u} h_{u,m,g,i}^*$. Using Wirtinger calculus and noting that $\bm{p}_m\in\mathbb{R}^3$ and $R$ is real-valued, the real gradient satisfies $\nabla_{\bm{p}_m} R = 2\Re \{ \sum_{u,g} \frac{\partial R}{\partial h^*} \frac{\partial h^*}{\partial \bm{p}_m} \}$. Exploiting $2\Re\{\mathrm{j}z\} = -2\Im\{z\}$, we obtain
\begin{equation}\label{equ.grad_pos_zf}
    \nabla_{\bm{p}_m} R = -\frac{4\pi}{\lambda} \Im \left\{ \sum_{g=1}^G \sum_{u=1}^U \xi_{u,m,g}^{(R)} \sum_{i=1}^{L_u} h_{u,m,g,i}^* \bm{k}_{\mathrm{tx},i,u} \right\}.
\end{equation}
The ZF spatial block updates the antenna positions by the projected-gradient step
\begin{equation}\label{equ.pos_update_zf}
    \bm{p}_m^{(n+1)} = \Pi_{\mathcal{S}_m} \left( \bm{p}_m^{(n)} + \eta_p \nabla_{\bm{p}_m} R \right),
\end{equation}
where $\eta_p > 0$ denotes a suitable position-update step size with $\bm{\Lambda}$ and the digital precoders fixed. The projection $\Pi_{\mathcal{S}_m}(\cdot)$ maps the updated point onto the feasible region~\eqref{equ.constraint} via component-wise clipping. Specifically, $[\Pi_{\mathcal{S}_m}(\bm{x})]_1 = 0$ enforces the planar constraint, while $[\Pi_{\mathcal{S}_m}(\bm{x})]_k = \min\!\big(\max([\bm{x}]_k,\, [\bar{\bm{p}}_m]_k - D_{\max}),\, [\bar{\bm{p}}_m]_k + D_{\max}\big)$ for $k \in \{2,3\}$ confines the displacement to the box region. The gradients of all antenna positions are evaluated at the current iterate, and one projected-gradient position trial is formed in each outer iteration. The feasible trial is retained if it does not decrease the current sum SE; otherwise, the current antenna positions are preserved. The overall ZF-based tri-domain procedure is summarized in Algorithm~\ref{alg:mm_zf_tridomain}.

\begin{algorithm}[!b]
    \footnotesize
    \KwIn{eCSI model; feasible sets $\{\mathcal{S}_m\}$; power budget $P_T$; $(N_{\max},\epsilon)$; step sizes.}
    \KwOut{Design $\{\bm{W}_g\}$, $\{\bm{\alpha}_m\}$, and $\{\bm{p}_m\}$.}
    Initialization: set $\bm{p}_m^{(0)}=\bar{\bm{p}}_m$ and choose feasible unit-norm $\bm{\alpha}_m^{(0)}$, $\forall m$\;
    Build $\bm{\Lambda}^{(0)}$ and $\{\bm{h}_{u,g}\}$ via~\eqref{equ.h_def}\;
    Digital initialization: obtain $\{\bm{W}_g^{(0)}\}$ via~\eqref{equ.O_g_def}-\eqref{equ.zf_normalize} and $R^{(0)}$ via~\eqref{equ.se}\;
    \For{$n=0$ \KwTo $N_{\max}-1$}{
        Channel update: evaluate $\{\bm{q}_{u,m,g}\}$ at $\{\bm{p}_m^{(n)}\}$ and form $\{\bm{h}_{u,g}\}$ via~\eqref{equ.h_def}\;
        Digital update: update $\{\bm{W}_g\}$ by normalized ZF via~\eqref{equ.O_g_def}-\eqref{equ.zf_normalize}\;
        EM update: update $\bm{\Lambda}$ via~\eqref{equ.oblique_cg_dir} and~\eqref{equ.lambda_update_zf}, and extract $\{\bm{\alpha}_m^{(n+1)}\}$\;
        EM refresh: rebuild $\{\bm{h}_{u,g}\}$ and update $\{\bm{W}_g\}$\;
        Position update: form a trial via~\eqref{equ.grad_pos_zf}-\eqref{equ.pos_update_zf}; accept it if the sum SE does not decrease and otherwise retain $\{\bm{p}_m^{(n)}\}$\;
        Position refresh: update $\{\bm{q}_{u,m,g}\}$, $\{\bm{h}_{u,g}\}$, and $\{\bm{W}_g\}$ at $\{\bm{p}_m^{(n+1)}\}$\;
        Objective update: evaluate $R^{(n+1)}$ via~\eqref{equ.se}\;
        \If{$\frac{|R^{(n+1)}-R^{(n)}|}{\max\{1,|R^{(n)}|\}}\le\epsilon$}{
            \textbf{break}\;
        }
    }
    \Return $\{\bm{W}_g\}$, $\{\bm{\alpha}_m\}$, and $\{\bm{p}_m\}$\;
    \caption{ZF-based tri-domain alternating optimization}
    \label{alg:mm_zf_tridomain}
    \LinesNumbered
\end{algorithm}
\section{Tri-Domain Alternating Optimization: WMMSE-Based Design}\label{S4}

Beyond the ZF-based baseline algorithm introduced in Section~\ref{S3}, we develop a tri-domain WMMSE-based algorithm to solve~\eqref{equ.problem} under a global total-power constraint in this section. Specifically, the WMMSE reformulation yields closed-form auxiliary and digital updates, while the EM and spatial blocks retain the update structure of Section~\ref{S3}.

{SEMRA-WMMSE retains the same EM and position variables and the same outer alternating structure. It replaces normalized ZF with auxiliary receiver, mean square error weight, and digital precoder updates that jointly account for interference and the total power budget. It is therefore the higher complexity extension of SEMRA-ZF rather than an independent design.}

\subsection{WMMSE Problem Transformation}\label{S4.1}

\subsubsection{Signal Estimation and Mean Square Error}
For each user-subcarrier pair, we introduce a scalar receive equalizer $\beta_{u,g} \in \mathbb{C}$ and the estimated symbol $\hat{s}_{u,g} \triangleq \beta_{u,g}^* y_{u,g}$. The mean square error (MSE) $\varepsilon_{u,g} \triangleq \mathbb{E}\{ |\beta_{u,g}^* y_{u,g} - s_{u,g}|^2 \}$ then expands via~\eqref{equ.rx} as
\begin{align}\label{equ.mse_explicit}
\begin{split}
    \varepsilon_{u,g} &= \underbrace{\left| \beta_{u,g}^* \bm{h}_{u,g}^H \bm{w}_{u,g} - 1 \right|^2}_{\text{signal distortion}} \\
    &\quad + \underbrace{\sum_{\substack{\ell=1 \\ \ell \neq u}}^{U} \left| \beta_{u,g}^* \bm{h}_{u,g}^H \bm{w}_{\ell,g} \right|^2}_{\text{multiuser interference}} + \underbrace{\sigma_n^2 |\beta_{u,g}|^2}_{\text{noise amplification}},
\end{split}
\end{align}
where $\bm{h}_{u,g}$ is defined in~\eqref{equ.h_def} and its dependence on $\bm{p}$ and $\bm{\alpha}_m$ is omitted for notational brevity.

\subsubsection{WMMSE Equivalence}
By the WMMSE equivalence~\cite{shi2011iteratively}, the sum-SE maximization problem~\eqref{equ.problem} can be equivalently reformulated by introducing auxiliary minimum mean square error (MMSE) receivers $\beta_{u,g}$ and positive MSE weights $\rho_{u,g}$ for each user-subcarrier pair, yielding
\begin{equation}\label{equ.wmmse_problem}
\begin{aligned}
    \mathcal{P}_{\text{WMMSE}}: \quad \min_{\mathcal{V}} \quad & F_{\mathrm{WMMSE}} \\
    \text{s.t.} \quad & \sum_{g=1}^{G} \sum_{u=1}^{U} \|\bm{w}_{u,g}\|_2^2 \le P_T, \\
    & \|\bm{\alpha}_m\|_2 = 1, \; \bm{\alpha}_m \in \mathbb{R}^K, \; \forall m, \\
    & \bm{p}_m \in \mathcal{S}_m, \; \forall m, \; \rho_{u,g} > 0, \; \forall u,g,
\end{aligned}
\end{equation}
where $\mathcal{V} \triangleq \{\{\bm{w}_{u,g}\}, \{\bm{\alpha}_m\}, \{\bm{p}_m\}, \{\beta_{u,g}\}, \{\rho_{u,g}\}\}$ denotes the collection of all optimization variables, and the WMMSE objective function is defined as
\begin{equation}\label{equ.wmmse_obj}
    F_{\mathrm{WMMSE}} \triangleq \sum_{g=1}^{G} \sum_{u=1}^{U} \left( \rho_{u,g} \varepsilon_{u,g} - \ln \rho_{u,g} \right).
\end{equation}

For fixed $\bm{W}$, $\bm{\alpha}_m$, and $\bm{p}$, optimizing the auxiliary variables $\beta_{u,g}$ and $\rho_{u,g}$ yields $\min_{\{\beta_{u,g}\},\,\{\rho_{u,g}>0\}} F_{\mathrm{WMMSE}} = UG - (\ln 2)\,R$. Hence, after eliminating the auxiliary variables, minimizing $F_{\mathrm{WMMSE}}$ over the remaining design variables is equivalent to maximizing $R$ in~\eqref{equ.se}. The reformulation also yields closed-form updates for $\beta_{u,g}$, $\rho_{u,g}$, and $\bm{w}_{u,g}$, whereas the EM- and spatial-domain blocks remain nonconvex and are handled in Sections~\ref{S4.3} and~\ref{S4.4}.

Substituting~\eqref{equ.mse_explicit} into~\eqref{equ.wmmse_obj} and expanding, the WMMSE objective becomes
\begin{equation}\label{equ.wmmse_obj_full}
\begin{split}
    F_{\mathrm{WMMSE}}
    = \sum_{g=1}^{G} \sum_{u=1}^{U} \Bigg[ \rho_{u,g} \Bigg( \left| \beta_{u,g}^* \bm{h}_{u,g}^H \bm{w}_{u,g} - 1 \right|^2 \\
    + \sum_{\substack{\ell=1 \\ \ell \neq u}}^{U} \left| \beta_{u,g}^* \bm{h}_{u,g}^H \bm{w}_{\ell,g} \right|^2 + \sigma_n^2 |\beta_{u,g}|^2 \Bigg) - \ln \rho_{u,g} \Bigg].
\end{split}
\end{equation}

\subsection{Digital Precoding and Auxiliary Variable Optimization}\label{S4.2}

With antenna positions $\bm{p}$ and EM-domain precoders $\bm{\alpha}_m$ fixed, the first block minimizes $F_{\mathrm{WMMSE}}$ over the auxiliary variables $\beta_{u,g}$, $\rho_{u,g}$, and the digital precoders $\bm{w}_{u,g}$ via BCD. Each subproblem is convex in its own variable block and therefore admits a closed-form update.

\subsubsection{Optimal Receive Equalizer Update}
We first optimize the scalar receive equalizer $\beta_{u,g}$ for each user-subcarrier pair while keeping $\bm{w}_{u,g}$ and $\rho_{u,g}$ fixed. Setting $\partial \varepsilon_{u,g}/\partial \beta_{u,g}^* = 0$ in~\eqref{equ.mse_explicit} yields the MMSE receiver given by
\begin{equation}\label{equ.beta_opt}
    \beta_{u,g}^{\star} = \frac{\bm{h}_{u,g}^H \bm{w}_{u,g}}{\sum_{\ell=1}^{U} |\bm{h}_{u,g}^H \bm{w}_{\ell,g}|^2 + \sigma_n^2}.
\end{equation}

\subsubsection{Optimal MSE Weight Update}
Setting $\partial(\rho_{u,g}\varepsilon_{u,g} - \ln\rho_{u,g})/\partial\rho_{u,g} = 0$ gives $\rho_{u,g}^{\star} = 1/\varepsilon_{u,g}$. When the optimal equalizer~\eqref{equ.beta_opt} is substituted, the minimum MSE reduces to $\varepsilon_{u,g}^{\min} = (1+\mathrm{SINR}_{u,g})^{-1}$ by the orthogonality principle, yielding
\begin{equation}\label{equ.rho_opt}
    \rho_{u,g}^{\star} = 1+\mathrm{SINR}_{u,g}.
\end{equation}

\subsubsection{Optimal Digital Precoder Update}
Finally, we optimize the digital precoders $\bm{w}_{u,g}$ subject to the total power constraint. Removing the terms independent of $\bm{w}_{u,g}$ from~\eqref{equ.wmmse_obj_full}, the subproblem becomes the following quadratically constrained quadratic program (QCQP)
\begin{equation}\label{equ.w_subproblem}
\begin{aligned}
    \min_{\{\bm{w}_{u,g}\}} \quad & \sum_{g=1}^G \sum_{u=1}^U \rho_{u,g} \varepsilon_{u,g} \\
    \text{s.t.} \quad & \sum_{g=1}^G \sum_{u=1}^U \|\bm{w}_{u,g}\|_2^2 \le P_T.
\end{aligned}
\end{equation}

To decouple the precoders across users, we first expand the weighted MSE. Interchanging the summation order over users $u$ and interferers $\ell$, the objective function can be rewritten as (with constants independent of $\bm{w}_{u,g}$ omitted)
\begin{equation}\label{equ.J_w}
    J(\{\bm{w}_{u,g}\}) = \sum_{g=1}^G \sum_{u=1}^U \left( \bm{w}_{u,g}^H \bm{\Psi}_g \bm{w}_{u,g} - 2\Re \left\{ \bm{w}_{u,g}^H \bm{z}_{u,g} \right\} \right),
\end{equation}
where we define the weighted channel covariance matrix as $\bm{\Psi}_g \triangleq \sum_{\ell=1}^U \rho_{\ell,g} |\beta_{\ell,g}|^2 \bm{h}_{\ell,g} \bm{h}_{\ell,g}^H \in \mathbb{C}^{M \times M}$ and the equivalent target vector as $\bm{z}_{u,g} \triangleq \rho_{u,g} \beta_{u,g} \bm{h}_{u,g} \in \mathbb{C}^{M}$.

We solve~\eqref{equ.w_subproblem} via the Lagrangian method. Let $\nu \ge 0$ denote the dual variable associated with the power constraint. The Karush-Kuhn-Tucker (KKT) stationarity condition $\partial \mathcal{L} / \partial \bm{w}_{u,g}^* = \bm{0}$ yields $(\bm{\Psi}_g + \nu \bm{I}_M) \bm{w}_{u,g} = \bm{z}_{u,g}$.
Since $\bm{\Psi}_g \succeq \bm{0}$, the matrix $\bm{\Psi}_g + \nu \bm{I}_M$ is positive definite for any $\nu > 0$, and the optimal precoder admits the closed-form solution
\begin{equation}\label{equ.w_opt}
    \bm{w}_{u,g}^{\star}(\nu) = (\bm{\Psi}_g + \nu \bm{I}_M)^{-1} \bm{z}_{u,g}.
\end{equation}
Since $\bm{\Psi}_g$ has rank at most $U$, it is singular whenever $U < M$. However, $\bm{z}_{u,g} = \rho_{u,g}\beta_{u,g}\bm{h}_{u,g}$ is collinear with $\bm{h}_{u,g}$, and the $u$-th term of $\bm{\Psi}_g$ is scaled by $\rho_{u,g}|\beta_{u,g}|^2$, so $\bm{z}_{u,g}$ lies in the column space $\mathrm{range}(\bm{\Psi}_g)$ of $\bm{\Psi}_g$ whenever $\beta_{u,g}\neq 0$, while the case $\beta_{u,g}=0$ gives $\bm{z}_{u,g}=\bm{0}$ trivially. Therefore, the limit $\lim_{\nu \to 0^+} \bm{w}_{u,g}^{\star}(\nu) = \bm{\Psi}_g^{\dagger} \bm{z}_{u,g}$ is well-defined, and we define $\bm{w}_{u,g}^{\star}(0) \triangleq \bm{\Psi}_g^{\dagger} \bm{z}_{u,g}$.
The optimal Lagrange multiplier $\nu^{\star}$ is determined by complementary slackness.
If $\sum_{g,u} \|\bm{w}_{u,g}^{\star}(0)\|_2^2 \le P_T$, then $\nu^{\star} = 0$ and the power constraint is inactive.
Otherwise, $\nu^{\star} > 0$ is the unique root of $\sum_{g,u} \|\bm{w}_{u,g}^{\star}(\nu)\|_2^2 = P_T$, which can be found by bisection since the left-hand side is monotonically decreasing in $\nu$.
Unlike the normalized ZF baseline in Section~\ref{S3}, this digital update enforces only the global total-power constraint and can therefore redistribute digital power across users and subcarriers.

\subsection{EM-Domain Radiation Pattern Optimization}\label{S4.3}

With $\bm{W}$, $\bm{p}$, $\beta_{u,g}$, and $\rho_{u,g}$ fixed, we optimize the EM-domain weights $\bm{\alpha}_m$, or equivalently the block-diagonal matrix $\bm{\Lambda}=\mathrm{Blkdiag}\{\bm{\alpha}_1,\ldots,\bm{\alpha}_M\}\in\mathbb{R}^{KM\times M}$. As in Section~\ref{S3.1zf}, we reuse the masked-oblique manifold framework while adapting the objective and its Euclidean gradient to the WMMSE formulation.

\subsubsection{Subproblem Formulation}
With $\bm{W}$, $\bm{p}$, $\beta_{u,g}$, and $\rho_{u,g}$ fixed, the EM-domain subproblem is
\begin{align}\label{equ.alpha_subproblem_lambda}
\begin{split}
    \min_{\bm{\Lambda}\in\mathcal{OB}} \quad
    & \mathfrak{f}_{\mathrm{EM}}(\bm{\Lambda})
    \triangleq \sum_{g=1}^{G}\sum_{u=1}^{U} \rho_{u,g}\varepsilon_{u,g} \\
    \text{s.t.}\quad
    & \bm{\Lambda} = \bm{\Lambda}\odot \bm{M}_0,
\end{split}
\end{align}
where $\varepsilon_{u,g}$ is given in~\eqref{equ.mse_explicit} and $\bm{M}_0$ is the block-diagonal mask from Section~\ref{S3.1zf}. Since the weights $\rho_{u,g}$ are fixed, the term $-\sum_{g,u}\ln\rho_{u,g}$ in~\eqref{equ.wmmse_obj} is an additive constant and is omitted.

\subsubsection{Euclidean Gradient of the EM Objective}
From~\eqref{equ.h_def}, the effective channel satisfies $\bm{h}_{u,g}^H=\bar{\bm{q}}_{u,g}^H\bm{\Lambda}$. Let $c_{u,\ell,g}\triangleq \bm{h}_{u,g}^H\bm{w}_{\ell,g}$ be defined as in~\eqref{equ.c_def}. We define the WMMSE MSE-residual kernel as
\begin{align}\label{equ.xi_def}
    \xi_{u,m,g}^{(\varepsilon)} &\triangleq \frac{\partial (\rho_{u,g} \varepsilon_{u,g})}{\partial h_{u,m,g}^*} \nonumber\\
    &= \rho_{u,g} \left( |\beta_{u,g}|^2 \sum_{\ell=1}^U c_{u,\ell,g}^* w_{\ell,m,g} - \beta_{u,g}^* w_{u,m,g} \right),
\end{align}
where $w_{\ell,m,g} = [\bm{w}_{\ell,g}]_m$. Stacking the kernel into $\bm{\xi}_{u,g}^{(\varepsilon)} \triangleq [\xi_{u,1,g}^{(\varepsilon)},\ldots,\xi_{u,M,g}^{(\varepsilon)}]^T \in \mathbb{C}^{M}$ and noting that $\bm{\Lambda}$ is real-valued, the chain rule yields
\begin{equation}\label{equ.grad_euc_wmmse_lambda}
    \nabla \mathfrak{f}_{\mathrm{EM}}(\bm{\Lambda})
    = \left( 2\Re\left\{ \sum_{g=1}^{G}\sum_{u=1}^{U} \bm{q}_{u,g}\big(\bm{\xi}_{u,g}^{(\varepsilon)}\big)^{T} \right\} \right)\odot \bm{M}_0.
\end{equation}

\subsubsection{Masked-Oblique Manifold Update}
With the Euclidean gradient in~\eqref{equ.grad_euc_wmmse_lambda}, we reuse the masked-oblique update~\eqref{equ.oblique_cg_dir}-\eqref{equ.lambda_update_zf} with $\mathfrak{f}(\cdot)$ replaced by $\mathfrak{f}_{\mathrm{EM}}(\cdot)$ in~\eqref{equ.alpha_subproblem_lambda}. The updated per-antenna weights $\bm{\alpha}_m$ are extracted from the diagonal blocks of $\bm{\Lambda}$ before the digital refresh in Algorithm~\ref{alg:wmmse_tridomain}. The initialization uses a maximum-ratio transmission (MRT) digital precoder followed by one global normalization to satisfy the total-power constraint at $n=0$.

\begin{algorithm}[!t]
    \footnotesize
    \KwIn{eCSI model; feasible sets $\{\mathcal{S}_m\}$; power budget $P_T$; $(N_{\max},\epsilon,I_{\mathrm{in}}^{\max}=15)$; step sizes.}
    \KwOut{Design $\{\bm{W}_g\}$, $\{\bm{\alpha}_m\}$, and $\{\bm{p}_m\}$.}
    Initialization: set $\bm{p}_m^{(0)}=\bar{\bm{p}}_m$ and choose feasible unit-norm $\bm{\alpha}_m^{(0)}$, $\forall m$\;
    Build $\bm{\Lambda}^{(0)}$ and $\{\bm{h}_{u,g}\}$ via~\eqref{equ.h_def}\;
    Digital initialization: obtain $\{\bm{W}_g^{(0)}\}$ by globally normalized MRT, update $\beta_{u,g}^{(0)}$ and $\rho_{u,g}^{(0)}$ via~\eqref{equ.beta_opt}-\eqref{equ.rho_opt}, and evaluate $F_{\mathrm{WMMSE}}^{(0)}$ via~\eqref{equ.wmmse_obj}\;
    \For{$n=0$ \KwTo $N_{\max}-1$}{
        Channel update: evaluate $\{\bm{q}_{u,m,g}\}$ at $\{\bm{p}_m^{(n)}\}$ and form $\{\bm{h}_{u,g}\}$ via~\eqref{equ.h_def}\;
        Digital-WMMSE update: update $\beta_{u,g}$, $\rho_{u,g}$, and $\{\bm{W}_g\}$ via~\eqref{equ.beta_opt}-\eqref{equ.w_opt} until the relative objective change is at most $\epsilon$ or $I_{\mathrm{in}}^{\max}$ sweeps\;
        EM update: update $\bm{\Lambda}$ via~\eqref{equ.oblique_cg_dir} and~\eqref{equ.lambda_update_zf}, and extract $\{\bm{\alpha}_m^{(n+1)}\}$\;
        Digital refresh: rebuild $\{\bm{h}_{u,g}\}$ and rerun the inner BCD under the same stopping rule\;
        Position update: form a trial via~\eqref{equ.grad_pos}-\eqref{equ.pos_update_wmmse}; accept it if the sum SE does not decrease and otherwise retain $\{\bm{p}_m^{(n)}\}$\;
        Objective update: update $\{\bm{q}_{u,m,g}\}$ and $\{\bm{h}_{u,g}\}$ at $\{\bm{p}_m^{(n+1)}\}$, rerun the inner BCD under the same stopping rule, and evaluate $F_{\mathrm{WMMSE}}^{(n+1)}$\;
        \If{$\frac{|F_{\mathrm{WMMSE}}^{(n+1)}-F_{\mathrm{WMMSE}}^{(n)}|}{\max\{1,|F_{\mathrm{WMMSE}}^{(n)}|\}}\le\epsilon$}{
            \textbf{break}\;
        }
    }
    \Return $\{\bm{W}_g\}$, $\{\bm{\alpha}_m\}$, and $\{\bm{p}_m\}$\;
    \caption{WMMSE-based tri-domain alternating optimization}
    \label{alg:wmmse_tridomain}
    \LinesNumbered
\end{algorithm}

\subsection{Spatial-Domain Antenna Position Optimization}\label{S4.4}

With $\bm{W}$, $\bm{\alpha}_m$, $\beta_{u,g}$, and $\rho_{u,g}$ fixed, the final block updates $\bm{p}\triangleq[\bm{p}_1^T,\ldots,\bm{p}_M^T]^T$ by minimizing $\mathfrak{f}_{\mathrm{pos}}(\bm{p}) \triangleq \sum_{g,u} \rho_{u,g}\varepsilon_{u,g}$ subject to $\bm{p}_m \in \mathcal{S}_m$, where the constant $-\sum_{g,u}\ln\rho_{u,g}$ is omitted as in~\eqref{equ.alpha_subproblem_lambda}.

\subsubsection{Spatial Gradient}
Repeating the derivation of Section~\ref{S3.2zf}, i.e., using the same decomposition in~\eqref{equ.h_decomp_zf}-\eqref{equ.v_def_zf} together with $\partial h_{u,m,g,i}^*/\partial \bm{p}_m = \mathrm{j}\frac{2\pi}{\lambda}\bm{k}_{\mathrm{tx},i,u} h_{u,m,g,i}^*$, but replacing $\xi_{u,m,g}^{(R)}$ by $\xi_{u,m,g}^{(\varepsilon)}$ from~\eqref{equ.xi_def}, gives
\begin{align}\label{equ.grad_pos}
    \bm{g}_m^{\text{pos}} &\triangleq \nabla_{\bm{p}_m} \mathfrak{f}_{\mathrm{pos}}(\bm{p}) \nonumber\\
    &= -\frac{4\pi}{\lambda} \Im \left\{ \sum_{g=1}^G \sum_{u=1}^U \xi_{u,m,g}^{(\varepsilon)} \sum_{i=1}^{L_u} h_{u,m,g,i}^* \bm{k}_{\mathrm{tx},i,u} \right\}.
\end{align}

\subsubsection{Position Update with Projection}
The spatial block forms one projected-gradient position trial in each outer iteration. The trial is computed over all antenna positions according to
\begin{equation}\label{equ.pos_update_wmmse}
    \bm{p}_m^{(n+1)} = \Pi_{\mathcal{S}_m} \left( \bm{p}_m^{(n)} - \eta_p \bm{g}_m^{\text{pos}} \right),
\end{equation}
which is the WMMSE counterpart of~\eqref{equ.pos_update_zf}. The feasible trial is retained if it does not decrease the current sum SE; otherwise, the current positions are preserved before the next outer tri-domain iteration.

\section{Low-Overhead Movement-Aided CE Scheme}\label{S5}

The tri-domain optimization developed in Sections~\ref{S3}-\ref{S4} requires a position-dependent eCSI model so that $\bm{q}_{u,m,g}$ can be evaluated for all users, antennas, and subcarriers at the current antenna positions. In the fixed-position EMRA baseline following~\cite{Ying2025TCOM}, the same parametric pipeline is applied on the reference array. One-dimensional (1D) ESPRIT is used for delay estimation, followed by two-dimensional (2D) ESPRIT for AoD estimation. Because the array geometry remains fixed, the angular resolution is fundamentally limited by the $M$ physical spatial samples. For a moderate array size, AoD estimation accuracy degrades and the resulting eCSI quality deteriorates.

We overcome this limitation by exploiting the spatial reconfigurability of SEMRA. At the uplink CE stage, a known training pattern is assigned to each of $N_{\mathrm{pat}}$ training blocks. Within every block, all $M$ antennas are simultaneously repositioned over $N_s$ successive position slots, with each antenna visiting the same $N_s$ prescribed observation positions. The resulting $N_sM$ distinct spatial samples form a densified virtual array while the $N_{\mathrm{pat}}$ blocks provide repeated observations under different known patterns. The proposed scheme adopts a parametric estimation framework in which the multipath delays, AoD angles, and an equivalent channel gain vector, all independent of the BS antenna positions, are extracted as intermediate parameters. This position-invariant parameterization decouples CE from the subsequent data-transmission design. During CE, the antennas move to observation positions for high-resolution parameter extraction, and the eCSI at the optimized positions is then assembled directly from the estimated parameters without additional pilot overhead.
Since uplink-downlink channel reciprocity holds in TDD, the BS-side angles estimated during the uplink CE phase coincide with the downlink AoD parameters $(\theta_{i,u}, \phi_{i,u})$ defined in Section~\ref{S2.2}, and we retain the downlink notation throughout this section.

\subsection{Virtual Array Construction}\label{S5.1}

\subsubsection{Observation Position Design}
For this specific CE procedure, we assume equal inter-element spacings $d \triangleq d_y = d_z$ and define $d_{\min} \triangleq d/4$. For each BS antenna~$m$, the reference position $\bar{\bm{p}}_m$ serves as the center around which $N_s = 4$ observation positions are defined as
\begin{equation}\label{equ.pos_schedule}
    \bm{p}_m^{(s)} = \bar{\bm{p}}_m + \bm{\Delta}_s, \quad s = 1, \ldots, N_s,
\end{equation}
with the displacement vectors
\begin{equation}\label{equ.delta_def}
    \bm{\Delta}_s = \begin{bmatrix} 0 \\ \psi_{y,s}\, d_{\min} \\ \psi_{z,s}\, d_{\min} \end{bmatrix},\;\;
    (\psi_{y,s}, \psi_{z,s}) \in \left\{\!\begin{aligned} &(+1,+1),\;(+1,-1),\\ &(-1,+1),\;(-1,-1) \end{aligned}\right\}\!,
\end{equation}
for $s = 1,\ldots,4$, respectively. Thus, each observation position is offset from $\bar{\bm{p}}_m$ by $\pm d_{\min}$ along both the $y$- and $z$-axes. The design requires $D_{\max} \ge d_{\min} = d/4$ so that all observation positions lie within the feasible region~\eqref{equ.constraint}. This is compatible with the non-overlapping condition $d \ge 2D_{\max} + D_{\mathrm{sep}}$ in Section~\ref{S2.4}. Both constraints are jointly satisfiable whenever $D_{\mathrm{sep}} < d/2$.

The CE stage uses a known set $\{\bm{\alpha}_{\mathrm{ce}}^{(t)}\}_{t=1}^{N_{\mathrm{pat}}}$ of real-valued pattern coefficient vectors in $\mathbb{R}^K$, where $\|\bm{\alpha}_{\mathrm{ce}}^{(t)}\|_2=1$. Pattern $\bm{\alpha}_{\mathrm{ce}}^{(t)}$ is held fixed across the $N_s$ position slots and all antennas within training block~$t$, so every block retains the coherent virtual-array steering used next. The set is used directly without CE-pattern optimization, while the different patterns provide additional observations for eCSI reconstruction.
We consider block-level operation in which all $N_{\mathrm{pat}}N_s$ pilot slots are completed within one quasi-static channel interval.

\begin{figure*}[!t]
	\centering
	\includegraphics[width=0.9\textwidth]{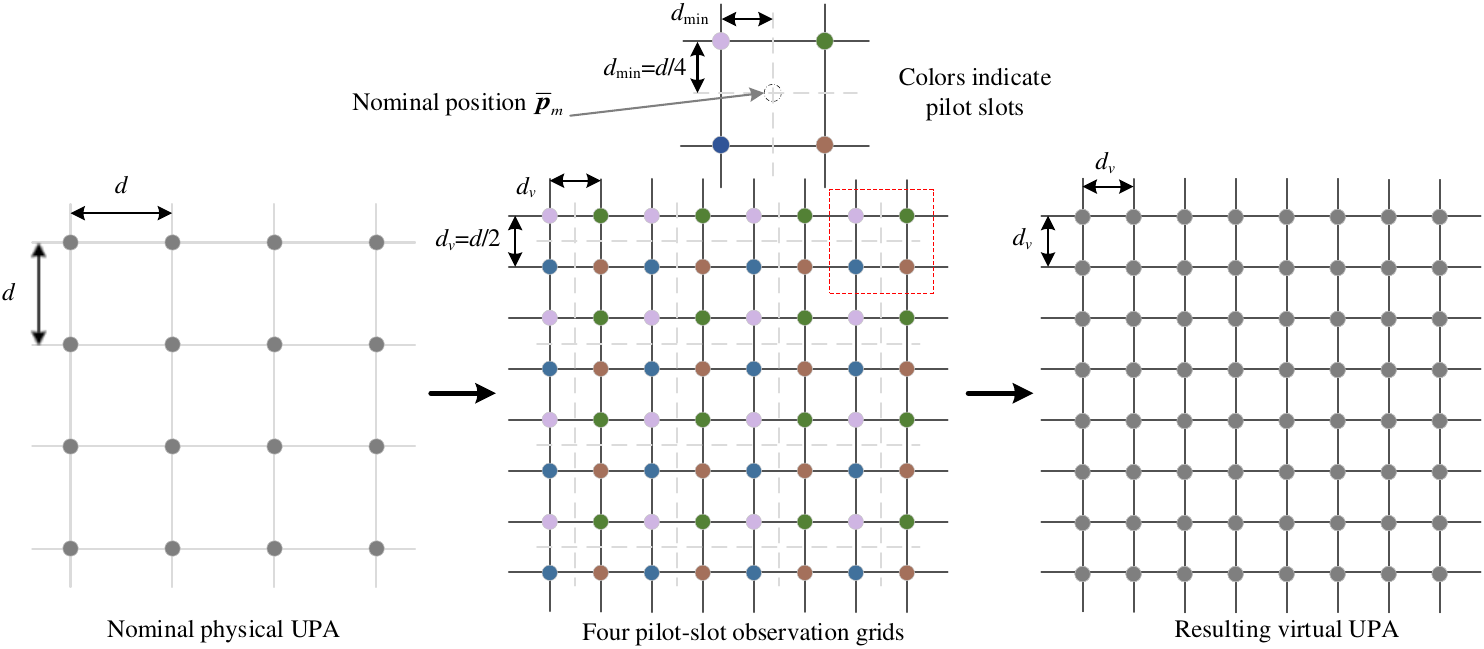}
	\caption{Virtual-array construction for movement-aided CE. Within each training block, the observation positions visited over $N_s=4$ position slots interleave an $M_y \times M_z$ reference UPA with spacing $d$ into a uniform $2M_y \times 2M_z$ virtual UPA with spacing $d_v=d/2$. The same position schedule is repeated for $N_{\mathrm{pat}}$ known training patterns. The illustration is shown for $M_y=M_z=4$.}
	\label{fig:virtual_upa_construction}
	\vspace*{-3mm}
\end{figure*}

\subsubsection{Virtual UPA Formation}
The resulting geometry is illustrated in Fig.~\ref{fig:virtual_upa_construction}. Within each training block, the four observation positions of antenna~$m$ form a $2\times 2$ arrangement centered at $\bar{\bm{p}}_m$ with spacing $2d_{\min}$, while the closest positions of adjacent antennas are separated by $d-2d_{\min}$. Setting $d_{\min}=d/4$ makes both spacings equal to $d/2$, so the $N_sM=4M$ samples form a uniform $2M_y\times2M_z$ virtual UPA with spacing $d_v=d/2$ and aperture $(M_y-\tfrac{1}{2})d\times(M_z-\tfrac{1}{2})d$. This array is used for 2D-ESPRIT AoD estimation, whereas delay estimation continues to rely mainly on frequency-domain shift invariance.

\subsection{Parametric Channel Estimation}\label{S5.2}

The discussion assumes the standard subspace-identifiability conditions for the minimum description length (MDL)/ESPRIT pipeline, under which the delay and spatial steering structures are resolvable for the chosen pilot and virtual-array dimensions, and we focus on a generic UE~$u$.

\subsubsection{Delay Estimation and sCSI Recovery}
In position slot~$s$ of training block~$t$, UE~$u$ transmits pilots on a dedicated subset $\mathcal{G}_u = \{g_{u,1}, \ldots, g_{u,J}\}$ of $J$ uniformly spaced subcarriers ($J \le G/U$), where the subsets of different users are disjoint to avoid mutual interference. The signal received at BS antenna~$m$ on pilot subcarrier~$g_j$ is
\begin{equation}\label{equ.ul_obs}
    y_{u,m,g_j}^{(t,s)} = \big(h_{u,m,g_j}^{(t,s)}\big)^{*}\, s_{u,g_j} + n_{u,m,g_j}^{(t,s)},
\end{equation}
where $s_{u,g_j}$ is the known pilot symbol and $n_{u,m,g_j}^{(t,s)}$ denotes receiver noise. Dividing by the pilot gives the effective observation $\tilde{y}_{u,m,g_j}^{(t,s)} = \big(h_{u,m,g_j}^{(t,s)}\big)^{*} + \tilde{n}_{u,m,g_j}^{(t,s)}$.
Under the downlink convention used here, reciprocity gives the uplink scalar coefficient as $\big(h_{u,m,g_j}^{(t,s)}\big)^*$.
The frequency-domain observations from all $N_{\mathrm{pat}}N_sM$ position-pattern pairs are divided by the known pilot symbols and stacked column-wise into a data matrix $\bm{Y}_d \in \mathbb{C}^{J \times N_{\mathrm{pat}}N_sM}$. Because the multipath channel~\eqref{equ.ch} is a sum of complex exponentials across frequency, $\bm{Y}_d$ can be written as $\bm{Y}_d = \bm{E}_d \bm{X}_d + \bm{N}_d$. Here, $\bm{E}_d \in \mathbb{C}^{J \times L_u}$ is a Vandermonde-structured delay steering matrix, $\bm{X}_d \in \mathbb{C}^{L_u \times N_{\mathrm{pat}}N_sM}$ collects the equivalent path gains across the position-pattern pairs, and $\bm{N}_d$ is the corresponding noise matrix.

The number of multipath components $L_u$ is estimated, rather than assumed known a priori, by applying the MDL criterion to the eigenvalues of the sample covariance matrix $\bm{Y}_d\bm{Y}_d^H \in \mathbb{C}^{J \times J}$~\cite{Wax1985Detection}, yielding the model-order estimate $\hat{L}_u$. The estimated order $\hat{L}_u$ is then used in the subsequent 1D-ESPRIT stage.

With $\hat{L}_u$ determined, the 1D-ESPRIT algorithm extracts the multipath delays $\{\hat{\tau}_{i,u}\}_{i=1}^{\hat{L}_u}$ from the shift-invariance structure of $\bm{E}_d$. The full-band sCSI $\hat{h}_{u,m,g}^{(t,s)}$ is then recovered by delay-domain least-squares (LS) fitting for every training block at the CE observation positions, i.e., the $M_y \times M_z$ reference-array positions for EMRA and the $N_sM$ positions of the $2M_y \times 2M_z$ virtual UPA for SEMRA.

\subsubsection{AoD Estimation on the Virtual Array}
At each observation position $\bm{p}_m^{(s)}$ in training block~$t$, the spatial-domain channel on subcarrier~$g$ can be expressed as
\begin{equation}\label{equ.ch_spatial}
    h_{u,m,g}^{(t,s)} = \sum_{i=1}^{L_u} \dot{x}_{i,u,g}^{(t)}\, e^{-\mathrm{j}\frac{2\pi}{\lambda}\bm{k}_{\mathrm{tx},i,u}^T \bm{p}_m^{(s)}},
\end{equation}
where $\bm{\omega}_{i,u} \triangleq [\omega_k(\theta_{i,u},\phi_{i,u})]_{k=1}^{K}$. The block-dependent composite gain is
\begin{equation}\label{equ.xdot_block}
    \begin{aligned}
    \dot{x}_{i,u,g}^{(t)}
    &\triangleq x_{i,u,g}\, f_{\mathrm{rx},u}(\vartheta_{i,u},\varphi_{i,u})
    \big((\bm{\alpha}_{\mathrm{ce}}^{(t)})^T \bm{\omega}_{i,u}\big)\\[-1mm]
    &\quad\times e^{-\mathrm{j}\frac{2\pi}{\lambda}\bm{k}_{\mathrm{rx},i,u}^T \bm{r}_u}.
    \end{aligned}
\end{equation}
Within training block~$t$, the radiation pattern $\bm{\alpha}_{\mathrm{ce}}^{(t)}$ is fixed across all position slots and antennas. Hence, the composite gain $\dot{x}_{i,u,g}^{(t)}$ does not depend on the antenna index~$m$ or the position-slot index~$s$, although it can change with~$t$.
For every training block, the $4M$ spatial observations therefore exhibit the same canonical steering-vector structure of a $2M_y \times 2M_z$ UPA with inter-element spacing $d_v$. The standard 2D-ESPRIT procedure~\cite{Liao2019TCOM} can consequently combine the training blocks after the re-indexing and spatial-smoothing steps described next.
Concretely, the virtual-array pilot observations are arranged into a data matrix of size $N_sM \times N_{\mathrm{pat}}J$, where rows index the $N_sM$ virtual-array positions and columns jointly index the training pattern and pilot subcarrier. The columns share the same spatial steering structure while their composite gains depend on the training pattern and subcarrier. The 2D-ESPRIT algorithm estimates the spatial frequencies $\{\mu_{i,u}, \nu_{i,u}\}$ from the sample covariance of this matrix after spatial smoothing.

Let $\{\hat{\mu}_{i,u}, \hat{\nu}_{i,u}\}_{i=1}^{\hat{L}_u}$ denote the spatial frequencies returned by 2D-ESPRIT, and define the virtual-array spatial frequency constant $\kappa_v \triangleq 2\pi d_v / \lambda$. Under the adopted virtual-array indexing, the spatial frequencies are related to the AoD by $\mu_{i,u} = \kappa_v \sin\theta_{i,u}\sin\phi_{i,u}$ and $\nu_{i,u} = -\kappa_v \cos\theta_{i,u}$. The minus sign in $\nu_{i,u}$ follows from the phase convention in~\eqref{equ.ch_spatial} together with the chosen indexing along the $z$-direction of the virtual UPA. We focus on BS-side departures in the front half-space, i.e., $\phi_{i,u}\in[-\pi/2,\pi/2]$. The AoD angles are then recovered on the principal branches as
\begin{equation}\label{equ.angle_recovery}
    \hat{\theta}_{i,u} = \arccos\!\left(-\frac{\hat{\nu}_{i,u}}{\kappa_v}\right), \;\; \hat{\phi}_{i,u} = \arcsin\!\left(\frac{\hat{\mu}_{i,u}}{\sqrt{\kappa_v^2 - \hat{\nu}_{i,u}^2}}\right),
\end{equation}
where $\hat{\theta}_{i,u}\in[0,\pi]$ and $\hat{\phi}_{i,u}\in[-\pi/2,\pi/2]$ by construction. Numerically, the arguments of $\arccos(\cdot)$ and $\arcsin(\cdot)$ are clipped to their feasible intervals before inversion. This mapping is one-to-one over the front half-space when $d_v \le \lambda/2$. For $d_v > \lambda/2$, we retain the same principal-branch inversion, while global uniqueness is not claimed.

With the estimated AoD pairs $\{(\hat{\theta}_{i,u}, \hat{\phi}_{i,u})\}$, the estimated direction vector is $\hat{\bm{k}}_{\mathrm{tx},i,u} = [\sin\hat{\theta}_{i,u}\cos\hat{\phi}_{i,u},\, \sin\hat{\theta}_{i,u}\sin\hat{\phi}_{i,u},\, \cos\hat{\theta}_{i,u}]^T$. The steering matrix $\hat{\bm{B}}_{u,m} \in \mathbb{C}^{\hat{L}_u \times \hat{L}_u}$ is then assembled at any desired position $\bm{p}_m$ via $[\hat{\bm{b}}_{u,m}]_i = e^{-\mathrm{j}\frac{2\pi}{\lambda}\hat{\bm{k}}_{\mathrm{tx},i,u}^T \bm{p}_m}$ (cf.\ Section~\ref{S2.2}), and the basis-function matrix $\hat{\bm{\Omega}}_u \in \mathbb{R}^{\hat{L}_u \times K}$ is constructed as $[\hat{\bm{\Omega}}_u]_{i,k} = \omega_k(\hat{\theta}_{i,u}, \hat{\phi}_{i,u})$ (cf.\ Section~\ref{S2.3}).

\subsubsection{Equivalent Channel Gain Estimation and eCSI Reconstruction}
Mirroring~\eqref{equ.v_def}, define $\tilde{\bm{\chi}}_{u,g} \in \mathbb{C}^{\hat{L}_u}$ as the equivalent path-gain vector under the estimated parametric model. The reconstructed eCSI then satisfies $\hat{\bm{q}}_{u,m,g}^H = \tilde{\bm{\chi}}_{u,g}^H \hat{\bm{B}}_{u,m} \hat{\bm{\Omega}}_u$, where only $\hat{\bm{B}}_{u,m}$ depends on $\bm{p}_m$.

Accordingly, each virtual observation satisfies
\begin{equation}\label{equ.obs_model}
    \hat{h}_{u,m,g}^{(t,s)*} = (\bm{\alpha}_{\mathrm{ce}}^{(t)})^T \hat{\bm{\Omega}}_u^T \hat{\bm{B}}_{u,m}^{(s)H} \tilde{\bm{\chi}}_{u,g}, \quad \forall m, s, t,
\end{equation}
where $\hat{\bm{B}}_{u,m}^{(s)}$ is evaluated at the observation position $\bm{p}_m^{(s)}$. Concatenating all $N_{\mathrm{pat}}N_s M$ observations into a single linear system gives $\hat{\bm{h}}_{u,g}^* = \bm{\Upsilon}_u \tilde{\bm{\chi}}_{u,g}$, with the observation vector $\hat{\bm{h}}_{u,g}^* \in \mathbb{C}^{N_{\mathrm{pat}}N_s M}$ and the measurement matrix
\begin{equation}\label{equ.Upsilon_def}
    \begin{aligned}
    \bm{\Upsilon}_u &\triangleq
    \begin{bmatrix}
    \bm{\Upsilon}_u^{(1)}\\[-1mm]
    \vdots\\[-1mm]
    \bm{\Upsilon}_u^{(N_{\mathrm{pat}})}
    \end{bmatrix}
    \in \mathbb{C}^{N_{\mathrm{pat}}N_s M \times \hat{L}_u},\\
    \bm{\Upsilon}_u^{(t)}
    &\triangleq \left[\hat{\bm{B}}_{u,1}^{(1)} \hat{\bm{\Omega}}_u \bm{\alpha}_{\mathrm{ce}}^{(t)},\, \ldots,\, \hat{\bm{B}}_{u,M}^{(N_s)} \hat{\bm{\Omega}}_u \bm{\alpha}_{\mathrm{ce}}^{(t)}\right]^H .
    \end{aligned}
\end{equation}

Provided that the columns of $\bm{\Upsilon}_u$ induced by the recovered paths are linearly independent, the matrix $\bm{\Upsilon}_u$ has full column rank. This requires nondegenerate steering responses over the $N_sM$ distinct observation positions, collectively nonzero responses from the training pattern set on every recovered path, and $N_{\mathrm{pat}}N_sM \ge \hat{L}_u$. Under this condition, the LS solution is
\begin{equation}\label{equ.v_ls}
    \hat{\bm{\chi}}_{u,g} = \left(\bm{\Upsilon}_u^H \bm{\Upsilon}_u\right)^{-1} \bm{\Upsilon}_u^H \hat{\bm{h}}_{u,g}^*, \quad \forall g.
\end{equation}
The eCSI at any target position $\bm{p}_m$ then follows from
\begin{equation}\label{equ.q_recon}
    \hat{\bm{q}}_{u,m,g}^H = \hat{\bm{\chi}}_{u,g}^H \hat{\bm{B}}_{u,m} \hat{\bm{\Omega}}_u, \quad \forall m, g,
\end{equation}
where $\hat{\bm{B}}_{u,m}$ is evaluated at the target position, either the reference position $\bar{\bm{p}}_m$ or the optimized position from Sections~\ref{S3}-\ref{S4}. Because only $\hat{\bm{B}}_{u,m}$ depends on $\bm{p}_m$, the CE-phase observation positions and the data-transmission positions need not coincide.

This CE procedure uses $N_{\mathrm{pat}}=3$ training blocks and $N_s=4$ position slots per block, with $J$ pilot subcarriers per slot per UE. The total per-user pilot overhead is therefore $N_{\mathrm{pat}}N_sJ=360$ symbols. For fair comparison, the EMRA baseline retains its fixed physical geometry over the same $N_{\mathrm{pat}}N_s=12$ slots and uses the same training pattern set and the same $J$ pilot subcarriers per slot, yielding the same total per-user pilot overhead.

Algorithm~\ref{alg:sera_ce} summarizes the virtual-array parameter estimation and target-position eCSI assembly.

\begin{algorithm}[!t]
    \footnotesize
    \KwIn{Pilot observations $\{y_{u,m,g_j}^{(t,s)}\}$; training patterns; pilot allocation; position schedule.}
    \KwOut{Position-dependent eCSI model.}
    Generate $\{\bm{p}_m^{(s)}\}$ via~\eqref{equ.pos_schedule}\;
    \For{$u=1$ \KwTo $U$}{
        Pilot stacking: form $\bm{Y}_d$ from the normalized observations\;
        Model order: estimate $\hat{L}_u$ via MDL on $\bm{Y}_d\bm{Y}_d^H$~\cite{Wax1985Detection}\;
        Delays: estimate $\{\hat{\tau}_{i,u}\}$ via 1D-ESPRIT\;
        sCSI: recover $\{\hat{h}_{u,m,g}^{(t,s)}\}$ on the virtual array by delay-domain LS\;
        Spatial frequencies: form the virtual-array snapshots and estimate $\{(\hat{\mu}_{i,u},\hat{\nu}_{i,u})\}$ via 2D-ESPRIT\;
        AoDs: obtain $\{(\hat{\theta}_{i,u},\hat{\phi}_{i,u})\}$ via~\eqref{equ.angle_recovery}\;
        Basis construction: build $\hat{\bm{\Omega}}_u$ and the observation-position steering matrices\;
        Equivalent gains: assemble $\bm{\Upsilon}_u$ and estimate $\hat{\bm{\chi}}_{u,g}$ via~\eqref{equ.v_ls}, $\forall g$\;
        Optional target assembly: obtain $\hat{\bm{q}}_{u,m,g}$ via~\eqref{equ.q_recon}\;
    }
    \Return position-dependent eCSI model\;
    \caption{Movement-aided parametric CE for SEMRA}
    \label{alg:sera_ce}
    \LinesNumbered
\end{algorithm}

\section{Simulation Results}\label{S6}

\subsection{Simulation Setup}\label{S6.1}

{Table~\ref{tab:simulation_parameters} consolidates the default simulation configuration. Unless otherwise stated, each parameter sweep changes only the quantity identified on its horizontal axis while retaining the remaining settings in the table.}

\begin{table}[!t]
	
	\caption{Default simulation parameters.}
	\label{tab:simulation_parameters}
	
	\centering
	\footnotesize
	\renewcommand{\arraystretch}{1.08}
	\begin{tabularx}{\columnwidth}{@{}lX@{}}
		\toprule
		{Parameter} & {Default value} \\
		\midrule
		{Carrier and waveform} & {$f_c=2.4$\,GHz, $\Delta f=30$\,kHz, $G=128$} \\
		{Reference array and users} & {$M_y=M_z=4$ $(M=16)$, $U=3$} \\
		{Radiation-basis size} & {$K=10^2$} \\
		{Multipath channel} & {$L_u=6$, $\sigma_\tau=100$\,ns} \\
		{Movement geometry} & {$d=\lambda/2$, $D_{\mathrm{sep}}=0.01\lambda$, $D_{\max}=0.245\lambda$} \\
		{Channel-estimation geometry} & {$N_s=4$, $d_v=\lambda/4$, $8\times8$ virtual array} \\
		{CE training schedule} & {$N_{\mathrm{pat}}=3$, $N_{\mathrm{pat}}N_s=12$ total slots} \\
		{Pilot allocation} & {$J=30$ subcarriers per user per slot} \\
		{Signal-to-noise ratio (SNR)} & {$\text{SNR}_\text{C}=\text{SNR}_\text{P}=20$\,dB, with $0$--$30$\,dB sweeps} \\
		{Outer iteration budget} & {$N_{\max}=10$} \\
		{Monte Carlo averaging} & {$1000$ independent channel realizations} \\
		\bottomrule
	\end{tabularx}
	\vspace*{-2mm}
\end{table}

We compare SEMRA-ZF and SEMRA-WMMSE with EMRA and with TFA/SMA baselines using downtilt, 3GPP~38.901, and isotropic patterns. EMRA and SEMRA are reported under both estimated and perfect eCSI, while TFA/SMA are reported under perfect sCSI.

{The evaluation focuses on communication-rate gains under a fixed transmit-power constraint. Actuator energy, movement latency, and calibration overhead are outside the present SE metric.}

\subsection{Channel Estimation Accuracy}\label{S6.2}

We evaluate CE with $\text{NMSE-S} \triangleq \frac{\sum_{u,m,g,t,s}|\hat{h}_{u,m,g}^{(t,s)}-h_{u,m,g}^{(t,s)}|^2}{\sum_{u,m,g,t,s}|h_{u,m,g}^{(t,s)}|^2}$ at each scheme's CE positions and $\text{NMSE-E} \triangleq \frac{\sum_{u,m,g}\|\hat{\bm{q}}_{u,m,g}-\bm{q}_{u,m,g}\|_2^2}{\sum_{u,m,g}\|\bm{q}_{u,m,g}\|_2^2}$ on a common reference array. Both ratios are computed per channel realization, converted to decibels, and then averaged. EMRA and SEMRA use the same per-user pilot overhead $N_{\mathrm{pat}}N_sJ=360$; orthogonal matching pursuit (OMP) benchmarks supplement the ESPRIT-based pipeline following~\cite{Ying2025TCOM}. NMSE-S reflects the delay/sCSI stage, whereas NMSE-E with estimated or oracle sCSI reflects eCSI reconstruction.

\begin{figure}[!t]
	\centering
	\includegraphics[width=1\columnwidth]{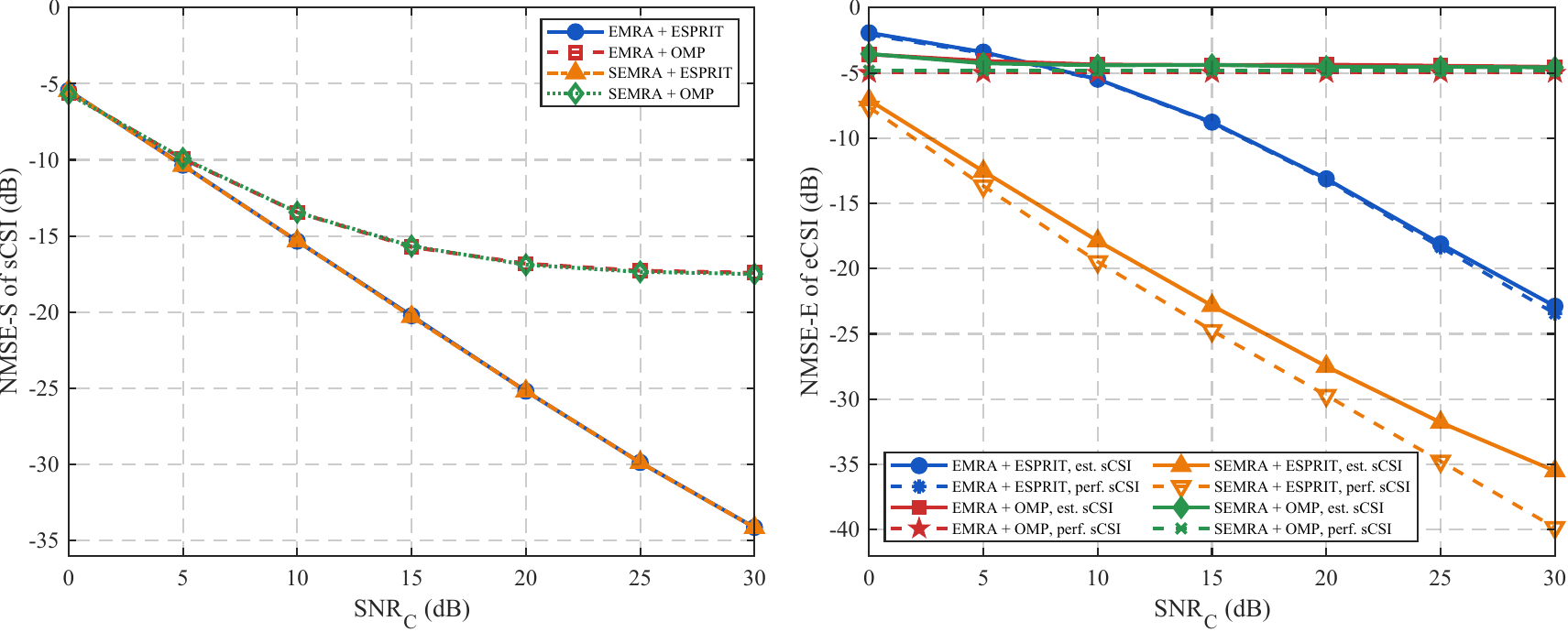}
	\caption{NMSE-S and NMSE-E versus $\text{SNR}_\text{C}$ at $d = \lambda/2$ for the EMRA baseline and the proposed SEMRA CE scheme.}
	\label{fig:nmse_vs_snr_c}
	\vspace*{-3mm}
\end{figure}

Fig.~\ref{fig:nmse_vs_snr_c} shows both metrics versus $\text{SNR}_\text{C}$ at $d=\lambda/2$. For NMSE-S, the EMRA and SEMRA curves almost overlap for each algorithm because this metric is dominated by frequency-domain delay estimation and LS-based sCSI recovery rather than by spatial sampling density. ESPRIT decreases steadily to about $-34.1$\,dB at $30$\,dB, whereas OMP saturates near $-17.5$\,dB due to its finite dictionary resolution. For NMSE-E with estimated sCSI, SEMRA+ESPRIT reaches about $-35.5$\,dB at $30$\,dB, compared with about $-22.9$\,dB for EMRA+ESPRIT. Their perfect-sCSI counterparts reach about $-39.8$\,dB and $-23.4$\,dB, respectively. In contrast, the OMP-based NMSE-E curves remain around $-4$ to $-5$\,dB at moderate and high SNR and show little separation between EMRA and SEMRA, indicating a dictionary-limited error floor. Thus, movement leaves sCSI recovery nearly unchanged, while the denser virtual array improves ESPRIT-based AoD estimation and eCSI reconstruction.

\begin{figure}[!t]
	\centering
	\includegraphics[width=1\columnwidth]{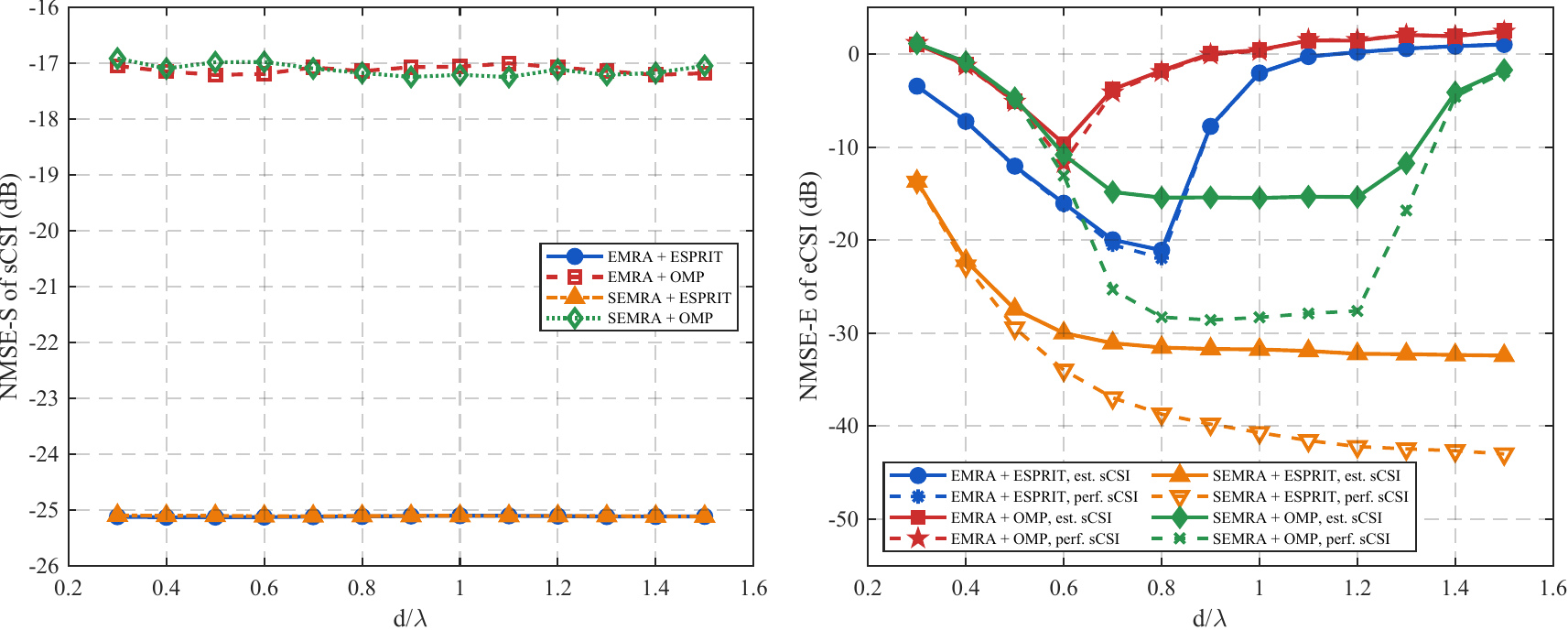}
	\caption{NMSE-S and NMSE-E versus inter-element spacing~$d/\lambda$ at $\text{SNR}_\text{C} = 20$\,dB for the EMRA baseline and the proposed SEMRA CE scheme.}
	\label{fig:nmse_vs_spacing}
	\vspace*{-3mm}
\end{figure}

Fig.~\ref{fig:nmse_vs_spacing} examines sensitivity to antenna spacing at $\text{SNR}_\text{C}=20$\,dB. The NMSE-S curves are nearly flat, with ESPRIT near $-25.1$\,dB and OMP near $-17.1$\,dB, because spacing mainly changes the spatial geometry used for AoD estimation and eCSI reconstruction rather than delay estimation and sCSI recovery. For NMSE-E, EMRA+ESPRIT initially benefits from the larger aperture and reaches a minimum of about $-21.1$\,dB at $d=0.8\lambda$ with estimated sCSI. Beyond this point, sparse physical sampling causes principal-branch ambiguity and grating lobes, and its NMSE-E rises sharply to about $1.0$\,dB at $d=1.5\lambda$. SEMRA+ESPRIT shows substantially greater robustness. It reaches about $-27.4$\,dB at $d=0.5\lambda$ and remains near $-32$\,dB from $0.9\lambda$ to $1.5\lambda$. Its perfect-sCSI curve further improves from about $-40$\,dB at $d=0.9\lambda$ to about $-43$\,dB at $d=1.5\lambda$. This advantage arises because the prescribed repositioning forms $N_sM$ distinct virtual spatial samples and observes them under $N_{\mathrm{pat}}$ known training patterns while retaining the same total pilot overhead as EMRA. The resulting denser observation geometry mitigates ambiguity at large spacing. SEMRA+OMP also outperforms EMRA+OMP from about $0.6\lambda$ onward, but degrades beyond $1.2\lambda$ because the finite dictionary remains sensitive to off-grid mismatch. Thus, movement-aided CE mainly improves eCSI reconstruction at large antenna spacings while leaving sCSI recovery nearly unchanged.

\subsection{Spectral Efficiency}\label{S6.3}

Unless otherwise stated, the SE plots compare EMRA, SEMRA-ZF, and SEMRA-WMMSE under perfect and estimated eCSI, with TFA/SMA baselines under perfect sCSI. EMRA versus SEMRA-ZF isolates the spatial-reconfiguration gain under normalized ZF, while SEMRA-WMMSE shows the additional digital-domain gain; estimated-eCSI curves use the ESPRIT-based CE pipeline and $\text{SNR}_\text{P}=20$\,dB by default.

\subsubsection{SE versus \texorpdfstring{$\text{SNR}_\text{C}$}{SNR\_C}}

\begin{figure}[!t]
	\centering
	\includegraphics[width=\columnwidth]{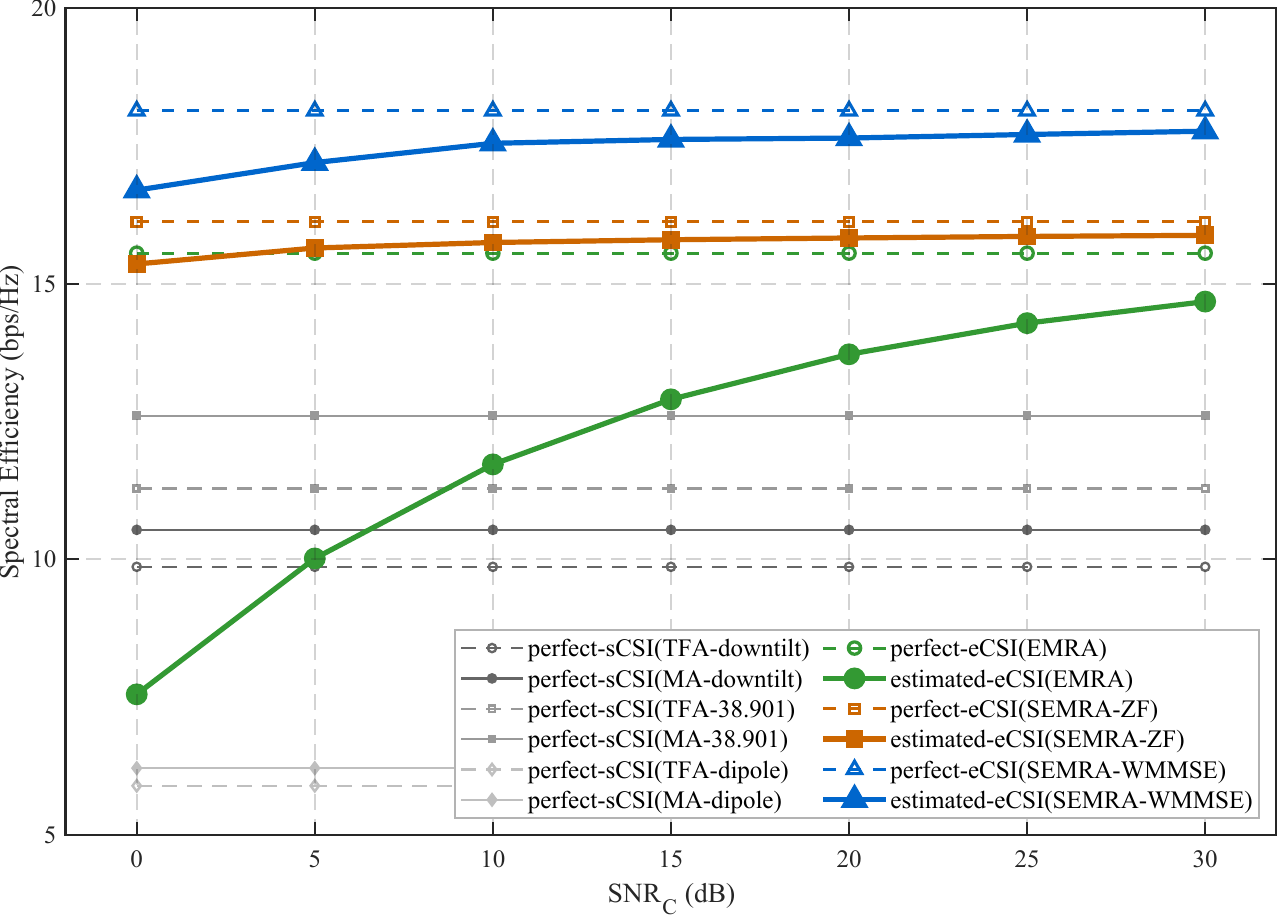}
	\caption{Sum SE versus $\text{SNR}_\text{C}$ when $\text{SNR}_\text{P} = 20$\,dB and $d = \lambda/2$.}
	\label{fig:se_vs_snr_c}
	\vspace*{-1mm}
\end{figure}

Fig.~\ref{fig:se_vs_snr_c} plots SE versus $\text{SNR}_\text{C}$. All estimated-eCSI curves increase with $\text{SNR}_\text{C}$ and approach their perfect-eCSI benchmarks, while the ordering SEMRA-WMMSE $>$ SEMRA-ZF $>$ EMRA is maintained over the whole range. The improvement is especially visible for EMRA, whose estimated-eCSI SE rises from the low-SNR regime but still remains below the SEMRA curves at high $\text{SNR}_\text{C}$. At $\text{SNR}_\text{C}=20$\,dB, the SEMRA perfect-versus-estimated gaps remain within about $0.50$\,bps/Hz, much smaller than the EMRA gap, which is consistent with the lower NMSE-E in Fig.~\ref{fig:nmse_vs_snr_c}. The remaining SEMRA-ZF gain under perfect eCSI confirms that spatial reconfiguration improves SE beyond CE accuracy alone, while SEMRA-WMMSE further benefits from stronger digital precoding.

\subsubsection{SE versus \texorpdfstring{$\text{SNR}_\text{P}$}{SNR\_P}}

\begin{figure}[!t]
	\centering
	\includegraphics[width=\columnwidth]{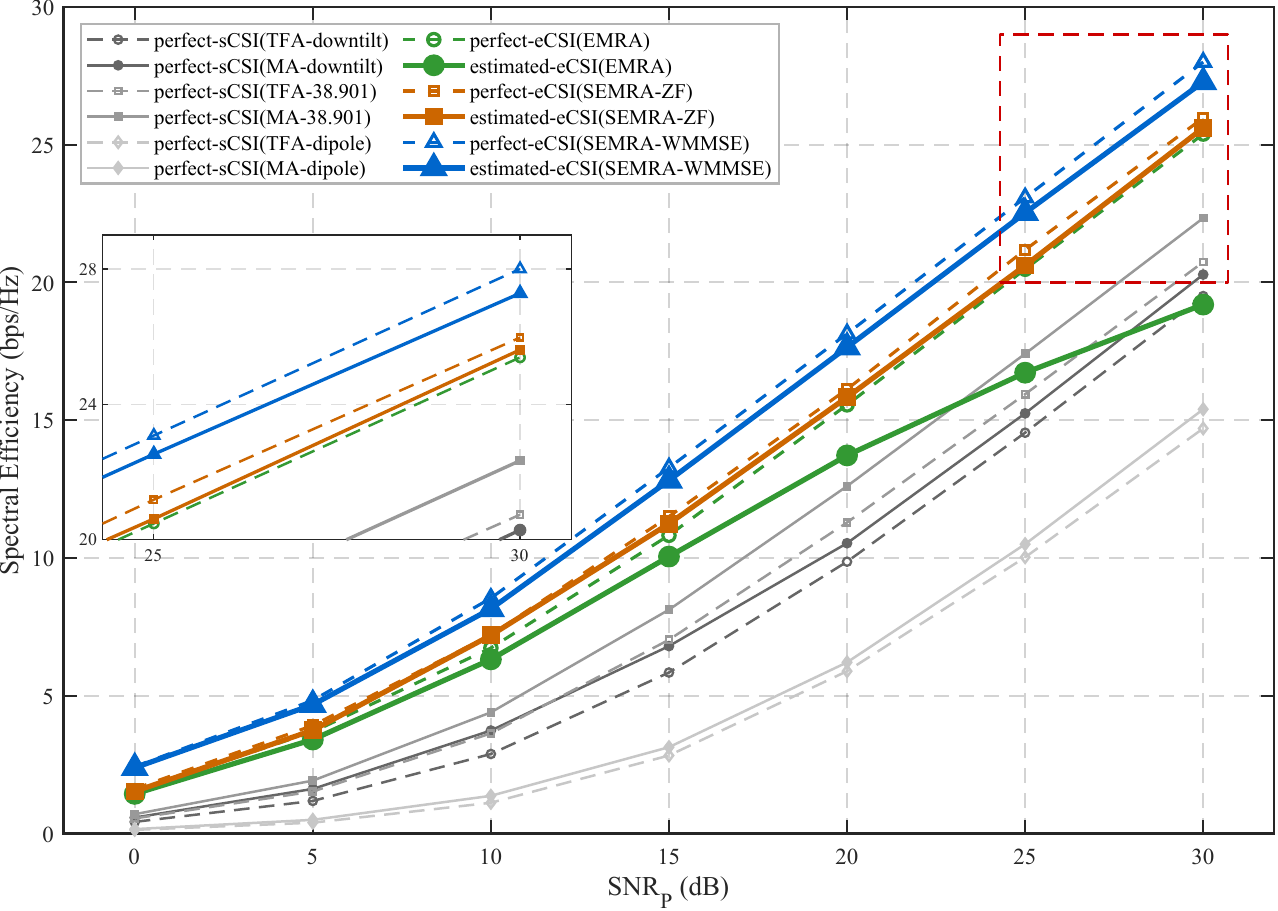}
	\caption{Sum SE versus $\text{SNR}_\text{P}$ when $\text{SNR}_\text{C} = 20$\,dB and $d = \lambda/2$.}
	\label{fig:se_vs_snr_p}
	\vspace*{-5mm}
\end{figure}

Fig.~\ref{fig:se_vs_snr_p} shows SE versus $\text{SNR}_\text{P}$ at $\text{SNR}_\text{C}=20$\,dB. All curves increase approximately linearly with $\text{SNR}_\text{P}$, and the SEMRA advantage over EMRA persists across the entire range. At the default $\text{SNR}_\text{P}=20$\,dB, SEMRA-WMMSE remains above EMRA and the fixed-pattern TFA/SMA baselines under estimated eCSI. The smaller perfect-versus-estimated gaps of the two SEMRA schemes again reflect their more accurate eCSI, while the persistent WMMSE-over-ZF gap of about $1.70$\,bps/Hz at high $\text{SNR}_\text{P}$ shows the benefit of the stronger digital design.

\subsubsection{SE versus Antenna Spacing}

\begin{figure}[!t]
	\centering
	\includegraphics[width=\columnwidth]{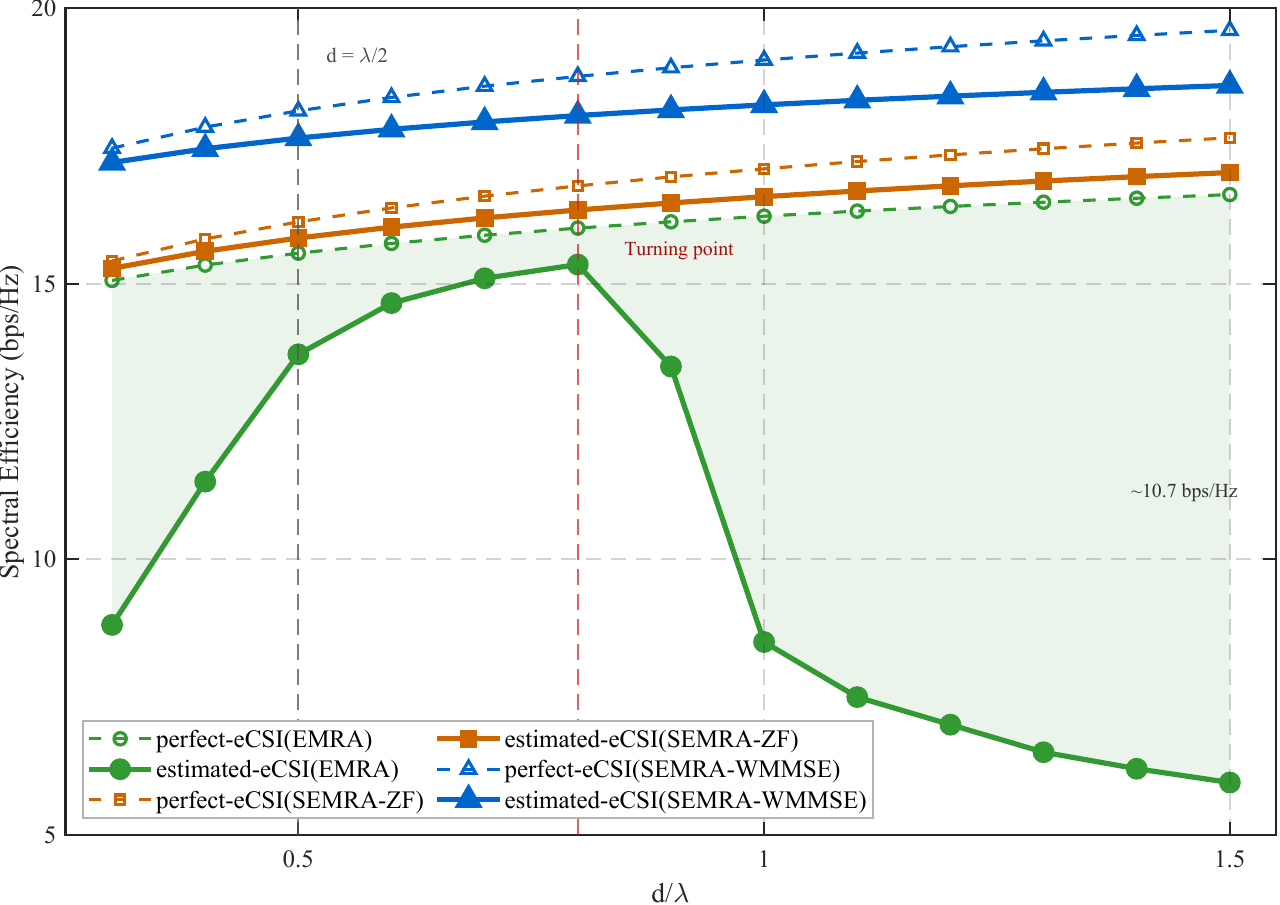}
	\caption{Sum SE versus inter-element spacing $d/\lambda$ when $\text{SNR}_\text{C} = 20$\,dB and $\text{SNR}_\text{P} = 20$\,dB.}
	\label{fig:se_vs_spacing}
	\vspace*{-5mm}
\end{figure}

Fig.~\ref{fig:se_vs_spacing} examines the impact of spacing when both SNRs are fixed at $20$\,dB. Varying $d$ jointly changes the physical aperture, CE-side aliasing behavior, and, for SEMRA, the feasible motion region. The perfect-eCSI curves increase with $d$, reflecting the aperture gain available without estimation errors. Under estimated eCSI, EMRA first benefits from aperture enlargement and peaks at about $15.40$\,bps/Hz around $d \approx 0.8\lambda$, but then collapses to about $6.00$\,bps/Hz at $d=1.5\lambda$, leaving the marked gap of about $10.70$\,bps/Hz to its perfect-eCSI counterpart. This SE turning point coincides with the NMSE-E minimum at $d=0.8\lambda$ in Fig.~\ref{fig:nmse_vs_spacing}, showing that the aperture benefit remains effective up to the point of the most accurate eCSI reconstruction. Once $d$ grows further, principal-branch ambiguity and grating lobes dominate, so the EMRA estimated-eCSI SE drops sharply despite the larger aperture. By contrast, SEMRA-ZF and SEMRA-WMMSE increase monotonically with $d$ and remain close to their perfect-eCSI benchmarks because SEMRA maintains low eCSI NMSE over the large-spacing region. At $d=1.5\lambda$, their estimated-eCSI SEs are about $17.00$ and $18.60$\,bps/Hz, respectively, confirming SEMRA's stronger robustness to enlarged inter-element spacings.

\subsubsection{SE versus Number of Users}

\begin{figure}[!t]
	\centering
	\includegraphics[width=\columnwidth]{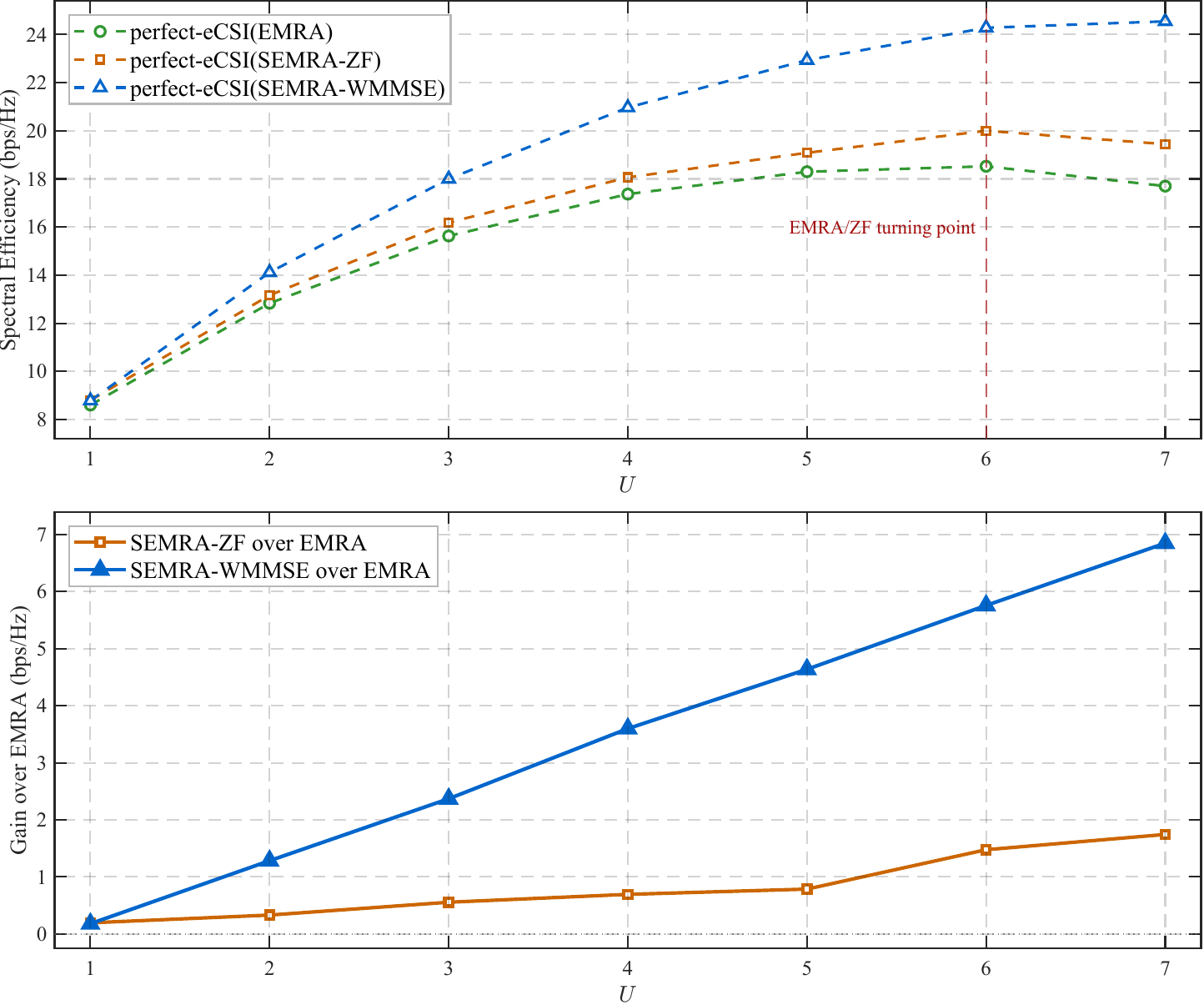}
	
	\caption{Sum SE and paired SEMRA gains over EMRA versus $U$ when $\text{SNR}_\text{P}=20$\,dB and $d=\lambda/2$ under perfect eCSI.}
	\label{fig:se_vs_users}
	
	\vspace*{-4mm}
\end{figure}

{Fig.~\ref{fig:se_vs_users} reports the sum SE and paired SEMRA gains over EMRA for $U=1,\ldots,7$ under the default physical and channel settings.}
{All three designs improve from the single-user case through moderate user loads, while both SEMRA variants retain positive paired gains over EMRA throughout the sweep.}
{At $U=4$ and $U=5$, the SEMRA-ZF gains over EMRA are $0.69$ and $0.79$\,bps/Hz, while the SEMRA-WMMSE gains are $3.60$ and $4.64$\,bps/Hz.}
{The lower panel removes the common variation in absolute SE and shows that the SEMRA-WMMSE gain becomes more pronounced as $U$ increases, which is consistent with the greater role of joint spatial, EM, and digital optimization under heavier multiuser interference.}
{SEMRA-ZF also remains above EMRA for every tested $U$, confirming a benefit from position optimization under the same normalized ZF design.}

\subsubsection{SE versus Maximum Antenna Displacement}

\begin{figure}[!t]
	\centering
	\includegraphics[width=\columnwidth]{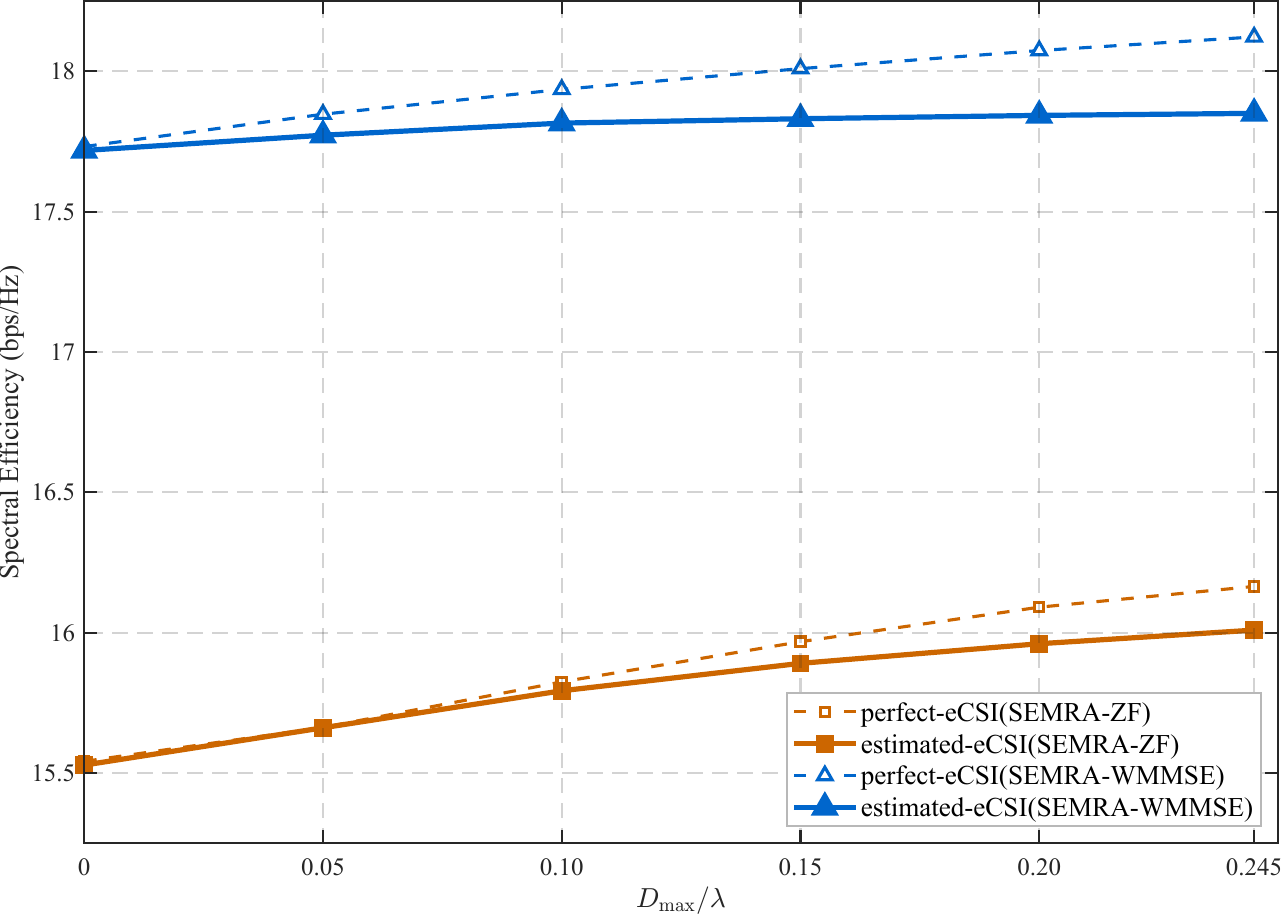}
	
	\caption{Sum SE versus $D_{\max}/\lambda$ when $d=\lambda/2$, $\text{SNR}_\text{C}=20$\,dB, and $\text{SNR}_\text{P}=20$\,dB.}
	\label{fig:se_vs_dmax}
	
	\vspace*{-5mm}
\end{figure}

{Fig.~\ref{fig:se_vs_dmax} reports monotonic SE increases across the feasible displacement range.}
{Since $d$ is held fixed, the changes isolate the value of additional position freedom from any enlargement of the reference aperture.}
{At $D_{\max}=0$, the antennas retain the reference geometry while the EM coefficients and digital precoders remain optimized.}
{Relative to this reference, the endpoint gains across the two designs and CSI conditions range from $0.13$ to $0.62$\,bps/Hz, and the perfect-eCSI curves remain above their estimated-eCSI counterparts.}
{Most of the endpoint improvement is already obtained at $D_{\max}=0.20\lambda$, so the curves begin to level before the full displacement allowance is reached.}
{With $d=\lambda/2$ and $D_{\mathrm{sep}}=0.01\lambda$, the collision-limited displacement bound is $0.245\lambda$. Both designs therefore benefit from the available local position freedom, and SEMRA-WMMSE yields the highest SE for every tested $D_{\max}$.}

\subsubsection{Convergence versus Outer Iteration Index}

\begin{figure}[!t]
	\centering
	\includegraphics[width=\columnwidth]{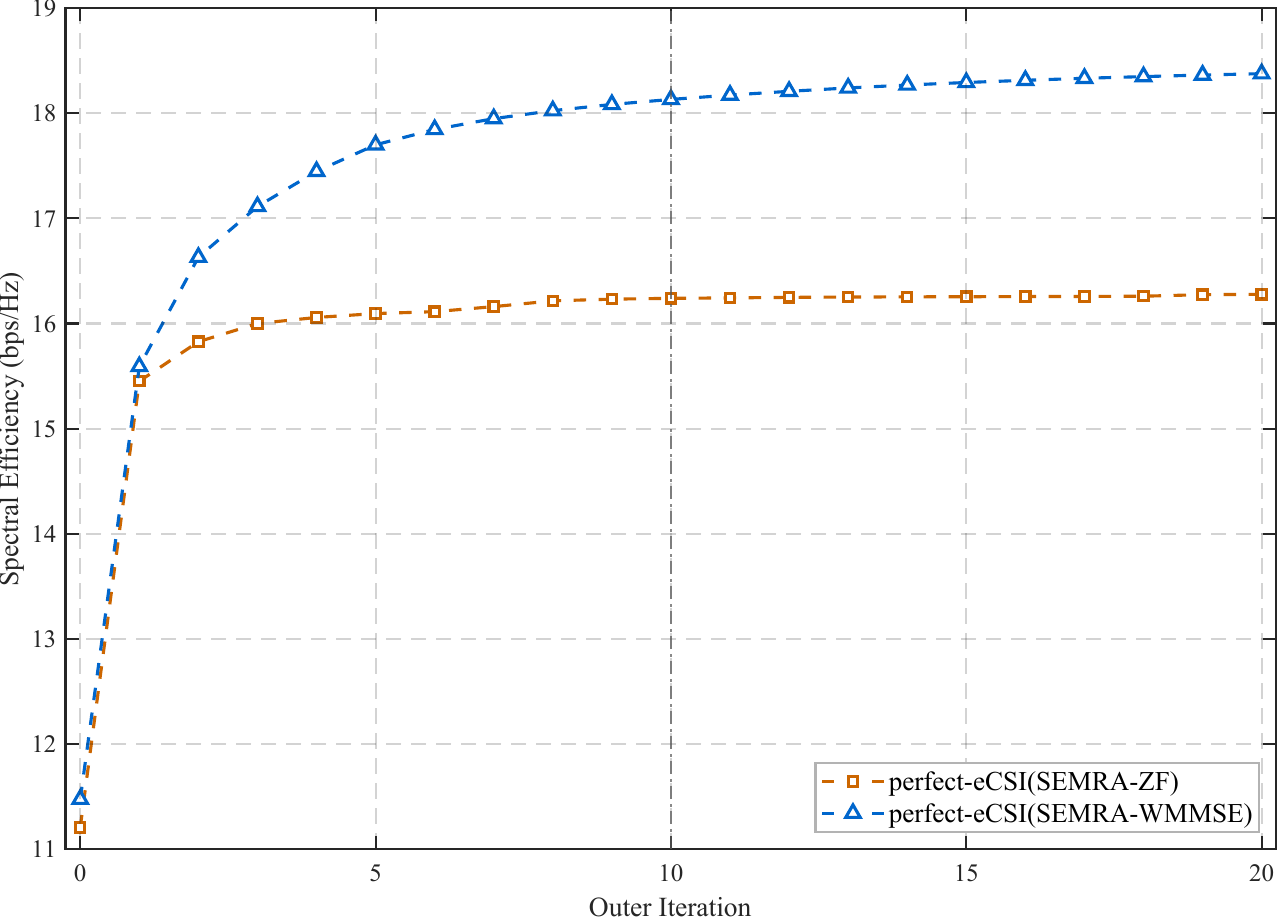}
	
	\caption{Sum SE versus the outer iteration index under perfect eCSI when $\text{SNR}_\text{P}=20$\,dB and $d=\lambda/2$. The default iteration budget is $N_{\max}=10$.}
	\label{fig:se_vs_iteration}
	
	\vspace*{-5mm}
\end{figure}

{Fig.~\ref{fig:se_vs_iteration} shows that the two algorithms obtain most of their observed improvement within the default iteration budget.}
{The first $10$ iterations capture $99.3\%$ and $96.4\%$ of the improvements observed by iteration $20$, leaving only $0.04$ and $0.25$\,bps/Hz of additional gain.}
{SEMRA-ZF plateaus earlier, while SEMRA-WMMSE continues refining the solution and approaches a higher SE more gradually.}
{These trends support $N_{\max}=10$ as a practical budget for both methods, while SEMRA-WMMSE remains above SEMRA-ZF throughout the common iteration horizon.}

\section{Conclusions}\label{S7}

{The proposed SEMRA framework combines movement-aided parametric CE with spatial, EM, and digital precoding for multiuser MIMO. During training, coordinated repositioning synthesizes a denser virtual array under the same per-user pilot overhead as EMRA, enabling AoD estimation and eCSI assembly at the transmission positions. SEMRA-ZF isolates the gain of spatial reconfiguration, while SEMRA-WMMSE further coordinates interference and power allocation. The simulations show improved eCSI accuracy and higher SE than EMRA, with larger gains at wider antenna spacings, positive paired gains across the tested user loads, and continued improvement throughout the tested displacement range. The convergence study further shows that the default ten iterations capture most of the improvement observed over twenty iterations and support the adopted ten-iteration budget. The study considers quasi-static geometric channels and block-level movement, and future work will examine measurement-based validation, online position control, and scalable optimization for larger systems.}

\bibliographystyle{IEEEtran}
\bibliography{refs}

\end{document}